\documentclass[times, 10pt]{article} 

\usepackage{mylatex8}
\usepackage{times}

\usepackage{amsmath}
\usepackage{amssymb}
\usepackage{latexsym} 
\usepackage{mytheorems}

\usepackage[all]{xy}
\usepackage{hyperref}

\input macros.tex

\title{The Variable Hierarchy for the Games  $\mu$-Calculus}
\author{Walid Belkhir  \and  Luigi Santocanale}
\begin{document}

\maketitle
\thispagestyle{empty}

%\abstract{ 
\begin{abstract}
  \sloppy Parity games are combinatorial representations of closed %\\
  Boolean \hbox{$\mu$-terms}. By adding to them draw positions, they
  have been organized by Arnold and one of the authors
  \cite{FOSSACS03,TCS333} into a $\mu$-calculus \cite{AN} whose
  standard interpretation is over the class of all complete lattices.
  % They introduced the usual composition, least and
  % greatest fixed-point operators. 
  As done by Berwanger et al.  \cite{BerwangerGraLen02,BerwangerLen05}
  for the propositional modal $\mu$-calculus, it is possible to
  classify parity games into levels of a hierarchy according to the
  number of fixed-point variables.  We ask whether this hierarchy
  collapses w.r.t. the standard interpretation.
  % of the games
  % $\mu$-calculus into the class of all complete lattices.
  We answer this question negatively by providing, for each $n \geq
  1$, a parity game $G_n$ with these properties: it unravels to a
  $\mu$-term built up with $n$ fixed-point variables, it is
  semantically equivalent to no game with strictly less than $n-2$
  fixed-point variables.
\end{abstract}
%}

%%% Local Variables: 
%%% mode: latex
%%% TeX-master: "0"
%%% End: 

\section{Introduction}

\newcommand{\refereesquestion}[1]{\marginpar{#1 ?}}
\newcommand{\mytitle}[1]{}

\mytitle{General context:  infinite two-player games and interactive computing, Berwanger's research ...}

% On aim is building a system that reacts correctly to its
% environment, and a winning strategy for a player in the game witnesses
% that the existence of a correct system.

%In computer science, a combination of infinite games, logic, and automata forms the theoretical basis for the synthesis and verification of reactive systems \cite{AVW:2003,Thomas97}.
% From proof theory point of view, formulae can be denoted by games and
% proofs (programs) can be denoted by winning strategies, providing a
% game semantics for programming languages \cite{Aabram94,blass}.
% Following this approach, a proof system for parity games was provided
% in \cite{Luigi}.

Recent work by Berwanger et al.
\cite{BerwangerThesis,berwanger,BerwangerGraLen02,BerwangerLen05}
proves that the expressive power of the modal $\mu$-calculus
\cite{kozen} increases with the number of fixed point variables. By
introducing the \emph{variable hierarchy} and showing that it does not
collapse, they manage to separate the $\mu$-calculus from dynamic game
logic \cite{Parikh}. Their work, solving a longstanding open problem,
may also be appreciated for %its new insights and
the new research paths\footnote{%
  We already pursued one of these paths in \cite{BelkSanto07}. We deal
  here with a problem of a more logical nature.}%
disclosed to the theory of fixed-points
% in computer science 
\cite{AN,bloomesik}.  The variable hierarchy may be defined for every
$\mu$-calculus and for iteration theories as well, since %just
one fixed-point operator is enough %sufficient
to define it. Thus, the question whether the variable hierarchy for a
$\mu$-calculus is strict is at least as fundamental as considering its
alternation-depth hierarchy.  In this paper we answer the question for
the \emph{games $\mu$-calculus over complete lattices}.

% Their ideas brought another insight in theory and suggest
% rich research paths.  Particularly they shown that the expressive
% power of propositional $\mu$-calculus increases in the number of fixed
% point variables. The core of their contributions is based on the
% classification of $\mu$-terms into hierarchy levels according to the
% number of fixed point variables rather than the traditional
% alternation depth classification \cite{clones,ANhier,Bradfield98}. The
% variable hierarchy classification has its merit at the conceptual
% level, it lead to separate dynamic game logic, DGL \cite{Parikh} from
% propositional $\mu$-calculus, by embedding DGL into the two-variable
% fragment of $\mu$-calculus. Such a separation can not be deduced, at
% least in a straitforward way, if we restrict ourselves to consider the
% alternation depth, since DGL captures all the alternation levels of
% $\mu$-calculus \cite{BerGamelogic03}.  In our opinion, the question
% concerning the strictness of the variable hierarchy can be asked for
% different $\mu$-calculi, and even for iteration theories
% \cite{bloomesik}, since just one fixed-point operator is considered.

Parity games are combinatorial representations of closed positive
Boolean $\mu$-terms. By adding to them draw positions (or free
variables), A. Arnold and L. Santocanale \cite{FOSSACS03,TCS333} have
structured parity games into \emph{the games $\mu$-calculus}.  In
other words, the authors defined substitution, least and greatest
fixed-point operators, as usual for $\mu$-calculi \cite{AN}. By
Tarski's theorem \cite{tarski} positive Boolean $\mu$-terms have a
natural interpretation in an arbitrary complete lattice. Such
interpretation transfers to a standard interpretation of this
$\mu$-calculus over the class of all complete lattices.\footnote{The
  interpretation in the class of distributive lattices makes the
  calculus trivial, since every $\mu$-term is equivalent to a term
  with no application of fixed-point operators.}
% In \cite{freemulat}, a syntactic
% preoder relation $\le$ was defined on the collection $\mathcal{G}$ of
% games. It was proved that the syntactic order relation characterize the semantic ones. 
The calculus, together with its canonical preorder, may also be
understood as a concrete description of the theory of binary infs and
sups, and of least and greatest fixed point over complete lattices,
what we called free $\mu$-lattices in \cite{freemulat}.

Let us recall the background of the games $\mu$-calculus.  The
interaction between two players in a game is a standard model of the
possible interactions between a system and its potentially adverse
environment.  Researchers from different communities are still working
on this model despite its introduction dates back at least fifteen
years \cite{Aabram94,Blass92,Nerode92} or more
\cite{blass72,joy77}.  It was proposed in \cite{Joyal97} to develop a
theory of communication grounded on similar game theoretic ideas and,
moreover, on algebraic concepts such as ``free lattice'' \cite{freese}
and ``free bicomplete category'' \cite{Joyal95}.  A first work pursued
this idea using tools of categorical logic \cite{cockettseely}. The
proposal was further developed in \cite{freemulat} where cycles were
added to lattice terms to enrich the model with possibly infinite
behaviors. As a result, lattice terms were replaced by positive
Boolean $\mu$-terms and their combinatorial representation, parity
games. The latter, one of the subtlest %most interesting
tool from the logics of programs, was introduced into the semantics of
computation. %programming languages.
Given two parity games $G,H$ the witness that the relation $G \leq H$
holds in every complete lattice interpretation is a winning strategy
for a prescribed player, Mediator, in a game $\homl G,H\homr$.  A game
$G$ may also be considered as modelling a synchronous communication
channel available to two users. Then, a winning strategy for Mediator
in $\homl G,H\homr$ witnesses the existence of an asynchronous
protocol allowing one user of $G$ to communicate with the other user
on $H$ ensuring absence of deadlocks.
Apart from its primary goal, that of describing complete lattices, a
major interest of this $\mu$-calculus stems from its neat
proof-theory, a peculiarity within the theory of fixed-point logics.
The idea that winning strategies for Mediator in the game $\homl G,H
\homr$ are sort of circular proofs was formalized in
\cite{Luigi}. More interestingly, proof theoretic ideas and tools --
the cut elimination procedure and $\eta$-expansion, in their game
theoretic disguise -- have proved quite powerful to solve deep
problems arising from fixed-point theory.  These are the
alternation-depth hierarchy problem \cite{TAC02} and the status of the
ambiguous classes \cite{FOSSACS03}.  In \cite{TCS333} the authors were
able to partially export these ideas to the modal $\mu$-calculus.  We
show here that similar tools success in establishing the strictness of
the variable hierarchy.

%  y combining the
% ideas relating the computational interactive systems to games
% \cite{Aabram94,McNaughton93,Nerode92} with the ideas relating games to
% free lattices and free bicomplete categories
% \cite{Joyal97,Joyal95,cockettseely}, it is convenient to model
% interactive computation by means of parity games -- free
% $\mu$-lattices in the sense of \cite{freemulat}-- and $\mu$-bicomplete
% categories.

\breath

\mytitle{The problem} %%
While dealing with  the variable hierarchy problem for the games $\mu$-calculus, we shall refer to two digraph complexity measures, the
\emph{entanglement} and the \emph{feedback}. 
% Their intuitive purpose
% is to measure to what extend cycles of a digraph are intertwined.  
The feedback of a vertex $v$ of a tree with back edges is the number
of ancestors of $v$ that are the target of a back edge whose source
is a descendant of $v$. The feedback of a tree with back edges is the
maximum feedback of its vertices.
% Its formal definition
% follows. 
The entanglement of a digraph $G$, denoted $\Ent{G}$, may be defined
as follows: \emph{it is the minimum feedback of its finite
  unravellings into a tree with back edges}. These measures are tied
to the logic as follows.  A $\mu$-term may be represented as a tree
with back-edges, the feedback of which corresponds to the minimum
number of fixed point variables needed in the $\mu$-term, up to
$\alpha$-conversion. Also, one may consider terms of a vectorial
$\mu$-calculus, i.e. systems of equations, and these roughly speaking
are graphs. The step that constructs a canonical solution of a system
of equations by means of $\mu$-terms amounts to the construction of a
finite unravelling of the graph. In view of these considerations,
asking whether a parity game $G$ is semantically equivalent to a
$\mu$-term with at most $n$-variables amounts to asking whether $G$
belongs to the level $\mathcal{L}_{n}$ defined as follows:
% As a matter of fact, one can consider a parity
% game as a vectorial representation of a system of equations, whereas
% the tree with back-edges, that unravels the game, represents a linear
% solution of the system given by means of $\mu$-term, the latter being
% computed at some variable.  \newline To make clear our objectives, let
% us define $\mathcal{L}_n$, for $n\ge 0$, the $n^{th}$ semantic level
% of variable hierarchy as follows.
\begin{equation}%
  \label{def:hierlevel}%
  \mathcal{L}_n=\set{G \in \mathcal{G}\; | \; G \sim H \textrm{ for some }H \in \mathcal{G}  \tst \Ent{H} \le n}%\,.
\end{equation}
Here $\mathcal{G}$ is the collection of parity games with draw
positions and $\sim$ denotes the semantic equivalence over complete
lattices.  In this paper we ask whether the variable hierarchy, made
up of the levels $\mathcal{L}_n$, collapses: is there a constant $k
\ge 0$, such that for all $n \ge k$, we have $\mathcal{L}_k =
\mathcal{L}_n$?  We answer this question negatively, there is no such
constant. We shall construct, for each $n \ge 1$, a parity game $G_n$
with two properties: (i) $G_n$ unravels to a tree with back edges of
feedback $n$, showing that $G_n$ belongs to $\mathcal{L}_n$, (ii)
$G_n$ is semantically equivalent to no game in $\mathcal{L}_{n-3}$.
Thus, we prove that the inclusions $\mathcal{L}_{n-3} \subseteq
\mathcal{L}_{n}$, $n \ge 3$, are strict.

\breath

\mytitle{Approach of the proof} %
% We combine in this work ideas of
% \cite{BerwangerGraLen02,BerwangerLen05} with the ideas of \cite{TAC02}
% on the alternation depth of the games $\mu$-calculus. On the one hand,
The games $G_n$ mimic the $n$-cliques of
\cite{BerwangerGraLen02,BerwangerLen05} that are shapes for hard
$\mu$-formulae built up with $n$ fixed point variables.
%  These
% $\mu$-formulae have been proved to be at least as \emph{entangled} as
% the $n$-cliques. 
This is only the starting point and, to carry on, we strengthen the
notion of \emph{synchronizing game}\footnote{%
  A synchronizing game has the property that there exists just one
  winning strategy for Mediator in $\langle G,G\rangle$, the copycat
  strategy.  } from \cite{TAC02} to the context of the variable
hierarchy. By playing with the $\eta$-expansion -- i.e. the copycat
strategy -- and the cut-elimination -- i.e. composition of strategies
-- we prove that the syntactical structure of a game $H$, which is
semantically equivalent to a \emph{strongly synchronizing} game $G$,
resembles that of $G$:
% a strongly synchronizing game $G$ imposes strong
% conditions on 
%  for a first approximation, $H$ must be at
% least as \emph{entangled} as $G_n$. The strong conditions may be
% precised as follows. 
every move (edge) in $G$ can be simulated by a non empty finite
sequence of moves (a path) of $H$; if two paths simulating distinct
edges do intersect, then the edges do intersect as well.  We formalize
such situation within the notion of \emph{$\star$-weak simulation}.
The main result is that if there is a $\star$-weak simulation of $G$
by $H$, then $\Ent{G} -2 \leq \Ent{H}$.
%
% This kind of weak simulation gives arise to the notion of
% $\star$-weak simulation, that establishes a useful connexion with
% entanglement of games as follows\footnote{Trop de ``as follows''
%   ...}.  Whenever there is a $\star$-weak simulation of $G_n$ by
% $H$, then the entanglement of $G_n$ is a lower bound for the
% entanglement of $H$, (up to a constant) for which the proof is not
% trivial.
The latter statement holds in the general context of digraphs, not
just for the games $\mu$-calculus, and might be of general use.

%  previous This result may
% be appreciated on its own, its scope does not restrict on parity
% games, and it can be of general use.

\mytitle{Perspectives and open questions}%%
We pinpoint next some aspects and open problems arising from the
present work. 
By combining the result on $\star$-weak simulations with the existence
of strongly synchronizing games $G_{n} \in \mathcal{L}_{n}$, we have
been able to prove that the inclusions
% The tools provided in our
% proof ensure that
$\mathcal{L}_{n-3} \subseteq \mathcal{L}_{n}$ are strict.  Yet we do not
know whether $\mathcal{L}_{n-1} \varsubsetneq \mathcal{L}_{n}$
and, at present, it is not clear that our methods 
can be
% so far
% developed
improved to establish the strictness of these
inclusions. % To prove the latter strict
% inclusions, we might try to reduce the constant $2$ in the
% $\star$-weak simulation Theorem \ref{weaksim:entag:prop}. 
We remark
by the way that we are exhibited with another difference with the
alternation hierarchy for which its infinity %of the hierarchy
implies that the inclusions between consecutive classes are strict.
Also, the reader will notice that the number of %\emph{
free variables %}
in the games $G_n$ increases with $n$. He might ask whether hard games
can be constructed using a fixed number of free variables. Here the
question is positively answered: most of the reasoning depends on free
variables forming an antichain so that we can exploit the fact that a
countable number of free variables (i.e. generators) can be simulated
within the free lattice on three generators \cite[\S 1.6]{freese}.
%%
%%
%  This would be done by proving that every
% game $G_n$ is a good representative for its equivalence class w.r.t
% variable hierarchy i.e.  for each $H \sim G_n$, we have $\Ent{G_n} \le
% \Ent{H}$. For the case of the alternation depth for both games
% $\mu$-calculus \cite{TAC02} and propositional $\mu$-calculus
% \cite{Bradfield98}, the authors proved the strictness of inclusion
% between any consecutive levels of alternation depth, showing that the
% nature of variable hierarchy problem differs intrinsically from the
% alternation depth problem.
%% 
% we think that is possible by substituting each free variable by an
% antichaine of a fixed number of free variables as done in
% \cite{Galvin61} to show that every free lattice of arbitrary set of
% generators can be embedded in a free lattice of a set of $3$
% generators.
Finally, the collection of parallel results on the modal
$\mu$-calculus and the games $\mu$-calculus -- compare for example
\cite{Bradfield98,TAC02} --
% Such kind of simulation
calls for the problem of relating these results by interpreting a
$\mu$-calculus into the another one.  While translations are a
classical topic in logic, we are not aware of results in this
direction for $\mu$-calculi.
% yet we are appealed by
% such a general question since
% translations might be of help in
% relating analogous and parallel results proved so far for different
% $\mu$-calculi.
 % Another problem has a more general appeal. A variety of analogous
% results have been proved for different $\mu$-calculi.  Yet, the proofs
% provided intrinsically differ and cannot be related in a
% straightforward way. A possible way to relate them would be to
% establish a suitable algebraic encoding or translation of a
% $\mu$-calculus into another one.

\mytitle{Organization of the paper}%
The paper is organized as follows. Section \ref{sec:paritygames}
introduces the necessary background on the algebra of parity games,
their organization into a $\mu$-calculus, their canonical preorder.
In section \ref{sec:entang}, we firstly recall the definition of
entanglement; then we define the $\star$-weak simulation between
graphs that allows to compare their entanglements. In section
\ref{sec:stronggames}, we define strongly synchronizing games and we
shall prove their \emph{hardness} w.r.t the variable hierarchy, in
particular every equivalent game to a strongly synchronizing one is
related with it by a $\star$-weak simulation.  In section
\ref{sec:ourgames}, we construct strongly synchronizing games of
arbitrary entanglement. We sum up the discussion in our main result,
Theorem \ref{main:theorem}.
% Because of lack of space, several proofs are
% given in the appendix.

\myparagraph{Notation, preliminary definitions and elementary
  facts.}%%
If $G$ is a graph, then a path in $G$ is a sequence of the form $\pi =
g_{0}g_{1}\ldots g_{n}$ such that $(g_{i},g_{i+1}) \in E_{G}$ for
$0\leq i < n$. A path is \emph{simple} if $g_{i} \neq g_{j}$ for $i,j
\in \set{0,\ldots ,n}$ and $i \neq j$. The integer $n$ is the length
of $\pi$, $g_{0}$ is the source of $\pi$, noted $\source \pi = g_{0}$,
and $g_{n} $ is the target of $\pi$, noted $\target \pi = g_{n}$.  We
denote by %$\Pi(G)$ the set of simple paths in $G$ and by
$\Pip(G)$ the set of simple non empty (i.e. of length greater than
$0$) paths in $G$. A pointed digraph $\langle V,E,v_{0} \rangle$ of
root $v_0$, is a \emph{tree} if for each $v \in V$ there exists a
unique path from $v_{0}$ to $v$.  A \emph{tree with back-edges} is a
tuple ${\cal T} = \langle V,T,v_{0},B \rangle$ such that $\langle
V,T,v_{0} \rangle$ is a tree, and $B \subseteq V \times V$ is a second
set of edges such that if $(x,y) \in B$ then $y$ is an ancestor of $x$
in the tree $\langle V,T,v_{0} \rangle$. We shall refer to edges in
$T$ as tree edges and to edges in $B$ as back edges. We say that $r
\in V$ is a return of ${\cal T}$ if there exists $x \in V$ such that
$(x,r) \in B$.  The \emph{feedback of a vertex} $v$ is the number of
returns $r$ on the path from $v_{0}$ to $v$ such that, for some
descendant $x$ of $v$, $(x,r)\in B$. The \emph{feedback of a tree with
  back edges} is the maximum feedback of its vertices.  We shall say
that a pointed directed graph $(V,E,v_{0})$ is a tree with back edges
if there is a partition of $E$ into two disjoint subsets $T,B$ such
that $\langle V,T,v_{0},B\rangle$ is a tree with back edges.

%% WALID EST QUE TU PENSES QUAND TU ECRIVES ???
%% C''EST QUOI CI DESSOUS ?
%  The feedback of a graph is the
% minimal feedback of its finite unravellings into a tree with
% back-edges.

%\paragraph{Simple paths in a tree with back edges.}
If ${\cal T}$ is a tree with back edges, then a path in ${\cal T}$ can
be factored as $\pi = \pi_{1} \ast \ldots \ast \pi_{n}\ast\tau$,
% \begin{align*}
%   \pi & = \pi_{1} \ast \ldots \ast \pi_{n}\ast\tau
% \end{align*}
where each factor $\pi_{i}$ is a sequence of tree edges followed by a
back edge, and $\tau$ does not contain back edges.
% The path $\tau$, which we shall refer to as the tail of $\pi$ or as
% $\tau(\pi)$, is a sequence (possibly empty) of tree edges.
Such factorization is uniquely determined by the occurrences of back
edges in $\pi$. For $i > 0$, let $r_{i}$ be the return at the end of
the factor $\pi_{i}$. Let also $r_{0}$ be the source of $\pi$.  Let
the $b$-length of $\pi$ be the number of back edges in $\pi$.
% $\pi_{i}$, 
i.e. $r_{i} = \target \pi_{i}$.
% For the $b$-length of
% $\pi$ we shall mean the integer $n$, that is, the number of back
% edges in $\pi$.
\begin{lemma}%
  \label{root:path}%
  \label{lemma:simplepathtwbe}%
  If $\pi$ is a simple path of
  $b$-length $n$, then $r_{n}$ is the vertex closest to the root
  visited by $\pi$. Hence, if a simple path $\pi$ lies in the subtree
  of its source, then it is a tree path.
\end{lemma}
We shall deal with trees with back-edges to which a given graph
unravels.
% that cover a given
% graph.
% thus we need the following definition.
\begin{definition}
  A \emph{cover} or \emph{unravelling} of a (finite) directed graph
  $H$ is a (finite) graph $K$ together with a surjective graph
  morphism $\rho : K \rTo H$ such that for each $v \in V_{K}$, the
  correspondence sending $k$ to $\rho(k)$ restricts to a bijection
  from $\set{ k \in V_{K} \mid (v,k) \in E_K}$ to $\set{ h \in V_{H}
    \mid (\rho(v),h) \in E_H}$.
% \begin{align*}
%     \rho(\set{x \in V_{K}\mid k \rightarrow x})
%     & = \set{y \in V_{H}\mid \rho(k) \rightarrow y}\,,
%   \end{align*}
%   and there is a bijection between $\set{(k,x) \in E_K}$ and
%   $\set{(\rho(k),\rho(x)) \in E_H}$ .
\end{definition}
The notion of cover of pointed digraphs is obtained from the previous
by replacing the surjectivity constraint by the condition that $\rho$
preserves the root of the pointed digraphs.

%%% Local Variables: 
%%% mode: latex
%%% TeX-master: "0"
%%% End: 

\newpage
\section{The Games $\mu$-Calculus} %
\label{sec:paritygames} 
In this section we recall the defintion of parity games with draws and
how they can be structured as a $\mu$-calculus. We shall skip the most
of the details and focus only on the syntactical preoder relation
$\le$ between $\mu$-terms that characterizes the semantical order
relation.

A \emph{parity game with draws} is a tuple $G = \homl
Pos_E^G,Pos_A^G,Pos_D^G, M^G, \rho^G \homr $ where:
\begin{myitemize}
\item $Pos_E^G,Pos_A^G,Pos_D^G$ are finite pairwise disjoint sets of
  positions (Eva's positions, Adam's positions, and draw positions),
\item $M^G$, the set of moves, is a subset of $(Pos_E^G\cup Pos_A^G)
  \times (Pos_E^G\cup Pos_A^G\cup Pos_D^G)$,
\item $\rho^G$ is a mapping from $(Pos_E^G\cup Pos_A^G)$ to $\N$. 
\end{myitemize}
Whenever an initial position is specified, these data define a game
between player Eva and player Adam. The outcome of a finite play is
determined according to the normal play condition: a player who cannot
move loses. It can also be a draw, if a position in $Pos_{D}^{G}$ is
reached.\footnote{Observe that there are no possible moves from a
  position in $Pos_{D}^{G}$.}  The outcome of an infinite play
$\set{\,(g_{k},g_{k + 1}) \in M^{G}\,}_{k \geq 0}$ is determined by
means of the rank function $\rho^{G}$ as follows: it is a win for Eva
iff the maximum of the set $\set{i \in \N \mid \exists \textrm{
    infinitely many } k
    %\\
    \textrm{ s.t. } 
    \rho^{G}(g_{k}) = i  \,}
$
is even.  To simplify the notation, we shall use $Pos_{E,A}^{G}$
for the set $Pos_{E}^{G} \cup Pos_{A}^{G}$ and use similar notations
such as $Pos_{E,D}^{G}$, etc. We let $Max^{G} = \max
\rho^{G}(Pos_{E,A}^{G})$ if the set $Pos_{E,A}^{G}$ is not empty, and
$Max^{G} = -1$ otherwise.

To obtain a $\mu$-calculus, as defined \cite[\S 2]{AN}, we label draw
positions with variables of a countable set $X$. If $\lambda^{G}:
Pos_{D}^{G} \rTo X$ is such a labelling and $p_{\star}^{G} \in
Pos_{E,A,D}^{G}$ is a specified initial position, then we refer to the
tuple $\homl G,p^{G}_{\star},\lambda^{G}\homr$ as a labeled parity
game.  We denote by $(G,g)$ the game that differs from $G$ only on the
starting position, i.e. $p_{\star}^{(G,g)}=g$, and similarly we write
$(G,g)$ to mean that the play has reached position $g$. We let
$\hat{x}$ be the game with just one final draw position of zero
priority and labeled with variable $x$.
%  We let $\chi_G=\set{x \in X \mid \exists p \in
%   Pos_{D}^G \tand \lambda^G(p)=x}$.
With $\mathcal{G}$ we shall denote the collection of all labeled
parity games; as no confusion will arise, we will call a labeled
parity game with simply ``game''.

As a $\mu$-calculus, formal composition and fixed-point operations may
be defined on ${\cal G}$; moreover, ${\cal G}$ has meet and join
operations.% The reader is invited to see \cite[\S 1, \S 2]{FOSSACS03}
%for the definitions of these operations. 
 When defining these operations on games we shall always assume that the sets of positions
of distinct games are pairwise disjoint.
\header{Meets and Joins.}  For any finite set $I$, $\Land[I]$ is
the game defined by letting $Pos_E=\emptyset$, $Pos_A= \set{p_0}$,
$Pos_D = I$, $M=\set{\,(p_0,i)\,\mid\, i \in I\,}$ (where $p_{0}
\not\in I$), $\rho(p_0)= 0$. The game $\Lor[I]$ is defined
similarly, exchanging $Pos_E$ and $Pos_A$.

\header{Composition Operation.}  Given two games $G$ and $H$ and a
mapping $\psi: P_{D}^{G} \rTo P_{E,A,D}^{H}$, the game $K = G
\circ_{\psi} H$ is defined as follows:
\begin{myitemize}
\item $Pos_E^K= Pos_E^G \cup  Pos_E^H$,
\item $Pos_A^K= Pos_A^G \cup Pos_A^H$,
\item $Pos_D^K=  Pos_D^H$,
\item $M^K= (M^G\cap (Pos_{E,A}^G \times Pos_{E,A}^G))
  \;\cup\; M^{H} 
  \\
  \mbox{\hspace{1mm}} \hfill  \;\cup\;
  \set{(p,\psi(p'))
    \mid (p,p')\in 
    % \hfill\mbox{\hspace{2mm}}\\
    % \mbox{\hspace{2mm}}\hfill
    M^G \cap
    (Pos_{E,A}^{G} \times Pos_{D}^{G}) }\,.
  $
\item $\rho^K$  is such that its restrictions to the positions of
  $G$ and $H$ are respectively equal to $\rho^G$ and $\rho^{H}$.
\end{myitemize}

\header{Sum Operation.} Given a finite collection of parity games
$G_{i}$, $i \in I$, their sum $H = \sum_{i \in I} G_{i}$ is defined in
the obvious way:
\begin{myitemize}
\item $P_{Z}^{H} = \bigcup_{i \in I} P_{Z}^{G_{i}}$, for $Z \in
  \set{E,A,D}$,
\item $M^{H} = \bigcup_{i \in I} M^{G_{i}}$, 
\item $\rho^H$  is such that its restriction to the positions of
  each $G_{i}$ is equal to $\rho^G_{i}$.
\end{myitemize}

\header{Fixed-Point Operations.}  If $G$ is a game, a system on $G$ is
a tuple $S = \langle E,A,M\rangle $ where:
\begin{myitemize}
\item $E$ and $A$ are pairwise disjoint subsets of $Pos_{D}^{G}$,
\item $M \subseteq (E \cup A)\times Pos_{E,A,D}^{G}$.
\end{myitemize}
Given a system $S$ and $\theta \in \set{\mu,\nu}$, we define the
parity game $\theta_{S}. G$:
\begin{myitemize}
\item $Pos_E^{\theta_{S}. G}= Pos_E^G \cup E$,
\item $Pos_A^{\theta_{S}. G}= Pos_A^G \cup A$,
\item $Pos_D^{\theta_{S}. G}= Pos_D^G - (E \cup A)$,
\item $M^{\theta_{S}. G}= M^G\cup M$,
\item $\rho^{\theta_{S}.G}$ is the extension of $\rho^G$ to $E \cup A$
  such that:
  \begin{myitemize}[-]
  \item if $\theta = \mu$, then $\rho^{\theta_{S}.G}$ takes on $E \cup
    A$ the constant value $Max^G$ if this number is odd or $Max^G +1$
    if $Max^G$ is even,
  \item if $\theta = \nu$, then $\rho^{\theta_{S}.G}$ takes on $E \cup
    A$ the constant value $Max^G$ if this number is even or $Max^G +1$
    if $Max^G$ is odd.
  \end{myitemize}
\end{myitemize}

\myparagraph{Semantics of $\mathcal{G}$.}%
The algebraic nature of parity games is better understood by defining
their semantics.  To this goal,
% Every parity game is constructible by means of the following
% operations, the reader is invited to see \cite[\S 1, \S 2]{FOSSACS03}
% for the full definitions. Meets and joins: $\Land_{I}$ and $\Lor_{I}$,
% the composition operation: $G\circ H$, the fixed point operation:
% $\theta_{S}.G$, where $\theta \in \set{\mu,\nu}$. To give the
% semantics of parity games
let us define the \textbf{predecessor game} $G^{-}$, for $G$ a game
such that $Max^{G} \neq -1$, i.e. there is at least one position in
$Pos_{E,A}^{G}$. Let $Top^{G} = \set{g \in
  Pos_{E,A}^{G}\mid\rho^{G}(g) = Max^{G}}$, then $G^{-}$ is defined as
follows:
\begin{myitemize}
\item $Pos^{G^{-}}_{E} = Pos^{G}_{E} - Top^{G}$, $Pos^{G^{-}}_{A} =
  Pos^{G}_{A} - Top^{G}$, $Pos^{G^{-}}_{D} = Pos^{G}_{D}\cup Top^{G}$,
\item $M^{G^{-}} = M^{G} - (Top^{G}\times Pos^{G}_{E,A,D})$,
\item $\rho^{G^{-}}$ is the restriction of $\rho^{G}$ to
  $Pos^{G^{-}}_{E,A}$.
\end{myitemize}
Given a complete lattice $L$, the interpretation of a parity game $G$
in $L$ is a monotone mapping of the form $ \eval{G}: L^{P^{G}_{D}}
\rTo L^{P^{G}_{E,A}}$. Here $L^{X}$ is the $X$-fold product lattice of
$L$ with itself so that, for $x \in X$, $\mathtt{pr}_{x} : L^{X} \rTo
L$ will denote the projection onto the $x$-coordinate.
% If $g \in Pos^{G}_{A,E}$ then
% $\eval{G_{g}}$ will denote the projection of $\eval{G}$ onto the $g$
% coordinate. Any parity game $G$ can be reconstructed in a unique way
% from the predecessor game $G^{-}$ by one application of some
% fixed-point operation $\theta_{S}$; moreover the predecessor game is
% ``simpler''.
The interpretation of a parity game is defined inductively.  If
$P^{G}_{E,A} = \emptyset$, then $L^{P^{G}_{E,A}} = L^{\emptyset} = 1$,
the complete lattices with just one element, and there is just one
possible definition of the mapping $\eval{G}$.  Otherwise, if
$Max^{G}$ is odd, then $\eval{G}$ is the parameterized least
fixed-point of the monotone mapping $ \; L^{P_{E,A}^{G}}\times
L^{P_{D}^{G}} \rTo L^{P_{E,A}^{G}} $ defined by the system of
equations:
\begin{align*}
  x_{g}
  & = 
  \left \{
    \begin{array}[c]{l@{\hspace{1.5mm}}l}
      \Lor[]\;
      \set{\,x_{g'}\,|\,(g,g') \in M^{G}\,}
      \,,
      \hspace{1.2mm}
      \textrm{if $g \in Pos_{E}^{G} \cap Top^{G}$}, \\[1mm]
      \Land[]\;
      \set{\,x_{g'}\,|\,(g,g') \in M^{G}\,}\,,
      \hspace{1.2mm}
      \textrm{if $g \in Pos_{A}^{G}\cap Top^{G}$}, \\[1mm]
      \mathtt{pr}_{g} \circ
      \eval{G^{-}}(X_{Top^{G}},X_{Pos^{G}_{D}})\,,
      \hspace{2mm}
      \textrm{otherwise}.
    \end{array}
    \right .
\end{align*}
If $Max^{G}$ is even, then $\eval{G}$ is the parameterized greatest
fixed-point of this mapping.

\myparagraph{The preorder on $\mathcal{G}$.}%
\label{sec:preorder}%
In order to describe a preorder on the class $\mathcal{G}$, we shall
define a new game $\homl G,H \homr$ for  a pair of games $G$ and $H$ in
$\mathcal{G}$. This is not a pointed parity game with draws as defined
in the previous section; to emphasize this fact, the two players will
be named Mediator and Opponents instead of Eva and Adam.

\begin{definition}
  The game $\homl G,H \homr$ is defined as follows:
  \begin{myitemize}
  \item The set of Mediator's positions is $ Pos_{A}^{G} \times
    Pos_{E,D}^{H} \; \cup \; Pos_{A,D}^{G} \times Pos_{E}^{H} \; \cup
    \; \mathcal{L}(M), $ and the set of Opponents' positions is $
    Pos_{E}^{G} \times Pos_{E,A,D}^{H} \; \cup \; Pos_{E,A,D}^{G}
    \times Pos_{A}^{H} \; \cup \; \mathcal{L}(O), $ where
    $\mathcal{L}(M), \mathcal{L}(O) \subseteq Pos_{D}^{G} \times
    Pos_{D}^{H}$ are the losing positions for Mediator and Opponents
    respectively. They are defined as follows. If $(g,h) \in
    Pos_{D}^{G}\times Pos_{D}^{H}$, then: if $\lambda^{G}(g) =
    \lambda^{H}(h)$, then the position $(g,h)$ belongs to Opponents,
    and there is no move from this position, hence this is a winning
    position for Mediator.  If $\lambda^{G}(g) \neq \lambda^{H}(h)$,
    then the position $(g,h)$ belongs to Mediator and there is no move
    from this position.  The latter is a win for Opponents.
    \item Moves of $\homl G,H \homr$ are either left moves $(g,h) \to
      (g',h)$, where $(g,g') \in M^G$, or right moves $(g,h) \to
      (g,h')$, where $(h,h') \in M^H$; however the Opponents can play
      only with Eva on $G$ or with $Adam$ on $H$.

\item A finite play is a loss for the player who can not move.  An
  infinite play $\gamma$ is a win for Mediator if and only if its left
  projection $\pi_G(\gamma)$ is a win for Adam, or its right
  projection $\pi_H(\gamma) $ is a win for Eva.
\end{myitemize}
\end{definition}

\begin{definition}
  If $G$ and $H$ belong to $\mathcal{G}$, then we declare that $G\leq
  H$ if and only if Mediator has a winning strategy in the game $\homl
  G,H \homr$ starting from position $(p_{\star}^{G},p_{\star}^{H})$.
\end{definition}
The following is the reason to consider such a syntactic relation:
\begin{theorem}[See \cite{freemulat}]
  The relation $\le$ is sound and complete with respect to the
  interpretation in any complete lattice, i.e. $G \le H$ if and only
  if $\eval G \le \eval H$ holds in every complete lattice.
\end{theorem}
In the sequel, we shall write $G \sim H$ to mean that $G \leq H$ and
$H \leq G$.  For other properties of the relation $\le$, see for
example Proposition $2.5$ of \cite{FOSSACS03}. One can prove that $G
\le G$, by exibing the \emph{copycat} strategy in the game $\homl G,G
\homr$: from a position $(g,g)$, it is Opponents' turn to move either
on the left or on the right board. When they stop moving, Mediator
will have the ability to copy all the moves played by the Opponents so
far from the other board until the play reaches the position
$(g',g')$.  There it was also proved that if $G \le H$ and $H \le K$
then $G\le K$, by describing a game $\homl G,H,K \homr$ with the
following properties: $(1)$ given two winning strategies $R$ on $\homl
G,H \homr$, and $S$ on $\homl H,K \homr$ there is a winning strategy
$R \Vert S$ on $\homl G,H,K \homr$, that is the composition of the
strategies $R$ and $S$, $(2)$ given a winning strategy $T$ on $\homl
G,H,K \homr$, there exists a winning strategy $T_{\backslash H}$ on
$\homl G,K \homr$.

The game $\homl G,H,K \homr$ is the fundamental tool that will allow
us to deduce the desired structural properties of games $H$ which are
equivalent to a specified game $G$, by considering the game $\homl
G,H,G \homr$, section \ref{sec:stronggames}. The game $\homl G,H,K
\homr$ is obtained by gluing the games $\homl G,H \homr$ and $\homl
H,K \homr$ on the central board $H$ as follows.
\begin{definition}
  Positions of the game  $\homl G,H,K \homr$ are triples $(g,h,k) \in Pos_{A,E,D}^{G} \times Pos_{A,E,D}^{H} \times Pos_{A,E,D}^{K}$ such that 
  \begin{myitemize}  
  \item the set of  Mediator's positions is  
    \begin{align*}
      &Pos_{A}^{G} \times Pos_{A,E,D}^{H} \times Pos_{E,D}^{K}
      %\\& 
      \; \cup \;Pos_{A,D}^{G} \times Pos_{A,E,D}^{H}
      \times
      Pos_{E}^{K}
      \; \cup \; \mathcal{L}(M)\,,
      % \end{align*}
    \intertext{and  the set of  Opponents' positions is}  
    %\begin{align*}
      &  Pos_{E}^{G} \times Pos_{A,E,D}^{H} \times Pos_{E,A,D}^{K} 
      %\\&
      \; \cup \;  
      Pos_{E,A,D}^{G} \times  Pos_{A,E,D}^{H} \times Pos_{A}^{K} 
      \cup \;  \mathcal{L}(O)\,,
    \end{align*}
    where $\mathcal{L}(M), \mathcal{L}(O) \subseteq Pos_{D}^{G} \times
    Pos_{A,E,D}^{H} \times Pos_{D}^{K}$ are positions of Mediator and
    Opponents, respectively, defined as follows. Whenever $(g,h,k) \in
    Pos_{D}^{G} \times Pos_{A,E,D}^{H} \times Pos_{D}^{K}$, then if $
    h \in Pos_{E,A}^H$, then the position $(g,h,k)$ belongs to
    Mediator, otherwise, i.e. $h\in Pos_{D}^H$, then the final
    position $(g,h,k)$ belongs to Opponents if and only if
    $\lambda^{G}(g)=\lambda^{H}(h)=\lambda^{K}(k)$.
  \item Moves of $\homl G,H,K \homr$ are either left moves $(g,h,k)
    \to (g',h,k)$ where $(g,g') \in M^{G}$ or central moves $(g,h,k)
    \to (g,h',k)$, where $(h,h') \in M^{H}$, or right moves $(g,h,k)
    \to (g,h,k')$, where $(k,k')\in M^{K}$; however the Opponents can
    play only with Eva on $G$ or with Adam on $K$.
  \item As usual, a finite play is a loss for the player who cannot
    move. An infinite play $\gamma$ is a win for Mediators if and only
    if $\pi_G(\gamma)$ is a win for Adam on $G$, or $\pi_K(\gamma)$ is
    a win for Eva on $K$.
\end{myitemize} 
\end{definition}

%%% Local Variables: 
%%% mode: latex
%%% TeX-master: "0"
%%% End: 

\newpage
\section{Entanglement and $\star$-Weak Simulations}%
\label{sec:entang}
%Towards the proof of the hierarchy theorem, we need to
Let us recall the main
tool which measures the combinatorial essence of the variable
hierarchy level on directed graphs.  This is the \emph{entanglement}
of a digraph $G$ and might already be defined as \emph{the minimum
  feedback of the finite unravelings of $G$ into a tree with back
  edges}.  
% As we have mentioned in the introduction, a $\mu$-term can
% be represented as a tree with back-edges, the feedback of which
% corresponds to the number of fixed point variables needed in the
% $\mu$-term up to $\alpha$-conversions.  
% Thus the feedback of a
% $\mu$-term $t$ gives already an interesting upper bound for the
% minimal number of fixed-point variables required in every term $t'$
% such that $\|t\|_{\mathcal{I}} =\|t'\|_{\mathcal{I}}$.
%% 
The entanglement of $G$ may also be characterized by means of a
special Robber and Cops game $\Ent{G,k}$, $k = 0,\ldots
,\card{V_{G}}$. This game, defined in \cite{berwanger}, is played by
Thief against Cops, a team\footnote{%
  We shall use the singular to emphasize that Cops constitute a team.
} %
of $k$ cops, as follows.
\begin{definition}
  The entanglement game $\Ent{G,k}$ of a digraph $G$ is defined by:
  \begin{myitemize}
  \item Its positions are of the form $(v,C,P)$, where $v \in V_{G}$,
    $C \subseteq V_{G}$ and $\card{C} \leq k$, $P \in \{Cops,
    Thief\}$.
  \item Initially Thief chooses $v_{0} \in V$ and moves to
    $(v_0,\emptyset,Cops)$.
  \item Cops can move  from $(v,C,Cops)$ to $(v,C',Thief)$
    where $C'$ can be
    \begin{myitemize}[-]
    \item $C$ : Cops skip,
    \item $C \cup\set{v}$ : Cops add a new Cop on the
      current position,
    \item $(C \setminus\set{x}) \cup \set{v}$ : Cops move a placed Cop
      to the current position.
    \end{myitemize}
  \item Thief can move from $(v,C,Thief)$ to $(v',C,Cops)$ if
    $(v,v') \in E_{G}$ and $v' \notin C$.
  \end{myitemize}
  Every finite play is a win for Cops, and every infinite play is a win
  for Thief. 
%   We let
%   \begin{align*}
%     \Ent{G} & = \min \set{ k\,|\,\text{Cops have a winning strategy in  $\Ent{G,k}$}}\,.
%   \end{align*}
%   The \emph{entanglement} of a digraph $G$, noted $\Ent{G}$, is the
%   minimum $k\geq 0$ such that Cops have a winning strategy in the
%   entanglement game $\Ent{G,k}$.
\end{definition}
The following will constitute our working definition of entanglement:
\emph{$\Ent{G}$, the entanglement of $G$, is the minimum $k \in \set{
    0,\ldots ,\card{V_{G}}}$ such that Cops have a winning strategy in
  $\Ent{G,k}$}.  The following proposition provides a useful variant
of entanglement games.
\begin{proposition} \label{modif:entag}
  Let $\ET{G,k}$ be the game played as the game $\Ent{G,k}$ apart
  that Cops is allowed to retire a number of cops placed on the
  graph. That is, Cops moves are of the form
  \begin{myitemize}
  \item $(g,C,Cops) \rightarrow (g,C',Thief)$ %%\\
    %%\myhspace{20mm}
    (generalized skip move),
  \item $(g,C,Cops) \rightarrow (g,C'\cup \set{g},Thief)$
   %  \\
%     \myhspace{30mm}
    (generalized replace move),
  \end{myitemize}
  where in both cases $C' \subseteq C$.
  Then Cops has a winning strategy in $\Ent{G,k}$ if and only if he
  has a winning strategy in $\ET{G,k}$.
\end{proposition}

%% ELSEWHERE
% Summarizing, \textbf{the variable hierarchy problem for the games
%   $\mu$-calculus} can be stated as follows: \emph{for every $n \ge
%   0$, is there a parity game $G_n$ of entanglement $n$ such that for
%   every game $H \sim G_n$, we have $\Ent{G} \le \Ent{H}$ ?}

%%% Local Variables: 
%%% mode: latex
%%% TeX-master: "0"
%%% End: 

\myparagraph{$\star$-Weak Simulations.} We define next a relation
between graphs, called \emph{$\star$-weak simulation}, to be used 
%% that we shall use 
to compare their entanglements.  Intuitively, there is a
$\star$-weak simulation of a graph $G$ by $H$ if every edge of $G$ is
simulated by a non empty finite path of $H$.  Moreover, two edges
$e_1,e_2$ of $G$ not sharing a common endpoint, are simulated by paths
$\pi_1, \pi_2$ that do not intersect.
% As a matter of fact,
% if $\pi_1$ and $\pi_2$ intersect, then $e_1$ and $e_1$ must be
% adjacent.
These simulations arise when considering games %$H$
which are semantically equivalent to %some
strongly synchronizing games, % $G$,
as defined in Section \ref{sec:stronggames}.

%% Remark : notation already in the introduction. 
% Before the formalization of these facts, we provide some notations.If $H$ is a graph, then a path in $H$ is a sequence of the form $\pi =
% h_{0}h_{1}\ldots h_{n}$ such that $(h_{i},h_{i+1}) \in E_{H}$ for
% $0\leq i < n$. A path is simple if $h_{i} \neq h_{j}$ for $i,j \in
% \set{0,\ldots ,n}$ and $i \neq j$. The integer $n$ is the length of
% $\pi$, $h_{0}$ is the source of $\pi$, %noted $\source \pi = h_{0}$,
% and $h_{n} $ is the target of $\pi$. % noted $\target \pi = h_{n}$.
% We denote

% Let $G,H$ be two graphs, and let $R \subseteq V_{G}\times V_{H}$. We
% say that $\pi \in \Pi(H)$ simulates $(g,g')$ w.r.t. $R$ if $g R
% \source \pi$ and $g' R \target \pi$. We shall write then $\pi
% \simulates[R] (g,g')$ or simply $\pi \simulates (g,g')$ if $R$ is
% understood.

\begin{definition}
  A \emph{weak simulation} $(R,\varsigma)$ of $G$ by $H$ is a binary
  relation $R \subseteq V_{G} \times V_{H}$ that comes with a partial
  function $\varsigma : V_G\times V_G \times V_H \rTo \Pi^{+}(H)$,
  such that:
  \begin{myitemize}
  \item $R$ is surjective, i.e. for every $g \in V_{G}$ there exists
    $h \in V_{H}$ such that $gR h$,
  \item $R$ is functional, i.e. if $g_{i} R h$ for $i = 1,2$, then $g_{1} = g_{2}$,
  \item if $gRh$ and $g \rightarrow g'$, then $\varsigma(g,g',h)$ is
    defined and $h' = \target\varsigma(g,g',h)$ is such that $g'Rh'$.
  \end{myitemize}
\end{definition}

Now we want to study conditions under which existence of a weak
simulation of $G$ by $H$ implies that $\Ent{G}$ is some lower bound
of $\Ent{H}$.
%  that is, we want to find a relation of the form $\Ent{G}
% \leq \Ent{H} + k$, for some fixed $k$.
To this goal, we abuse of notation and write $h \in
\varsigma(g,g',h_{0})$ if $\varsigma(g,g',h_{0}) = h_{0}h_{1}\ldots
h_{n}$ and, for some $i \in \set{0,\ldots ,n}$, we have $h = h_{i}$.
If $G = (V_{G},E_{G})$ is a directed graph then its undirected version
$S(G) = (V_{G},E_{S(G)})$ is the undirected graph such that
$\couple{g,g'} \in E_{S(G)}$ iff $(g,g') \in E_{G}$ or $(g',g) \in
E_{G}$. Thus we say that $G$ has \emph{girth at least $k$} if the
shortest cycle in $S(G)$ has length at least $k$, $G$ does not contain
loops, and $(g,g') \in E_{G}$ implies $(g',g) \not\in E_{G}$.
\begin{definition}\label{def:starwek}
  We say that a weak simulation $(R,\varsigma)$ of $G$ by $H$ is a
  $\star$-weak simulation (or that it has the $\star$-property) if $G$
  has girth at least $4$, and if 
  $(g,g'),(\tilde{g},\tilde{g}')$ are distinct edges of $G$ and $h \in
  \varsigma(g,g',h_{0}),\varsigma(\tilde{g},\tilde{g}',\tilde{h}_{0})$,
  then $\card{\set{g,g',\tilde{g},\tilde{g}'}} = 3$.
\end{definition}
We explain next this property. Given $(R,\varsigma)$, consider 
% the set
\begin{align*}
  C(h) & = \set{(g,g')\in E_G \mid \exists h_{0} \tst h \in \varsigma(g,g',h_{0}) }\,.
\end{align*}
\begin{lemma} \label{LEMMA:uniq:center} Let $(R,\varsigma)$ be a
  $\star$-weak simulation of $G$ by $H$.  If $C(h)$ is not empty, then
  there exists an element $c(h) \in V_{G}$ such that for each $(g,g')
  \in C(h)$ either $c(h) = g$ or $c(h) = g'$. If moreover $\card{C(h)}
  \geq 2$, then this element is unique.
\end{lemma}
That is, $C(h)$ considered as an undirected graph, is a star. Since
$c(h)$ is unique whenever $\card{C(h)} \geq 2$, then $c(h)$ is a
partial function which is defined for all $h$ with $\card{C(h)} \ge 2$.
This allows to define a partial function $f : V_{H} \rTo V_{G}$, which
is defined for every $h$ for which $C(h) \neq \emptyset$, as
follows:
\begin{align}
  \label{eq:deff}
  f(h) & = 
  \begin{cases}
    c(h), & \card{C(h)}\geq 2\,,  \\
    g, & \text{if } C(h) = \set{(g,g')} \tand 
    % \\
%     & \hspace{10mm} 
    h \text{ has no
      predecessor in } H\,,
    \\
    g',& 
     \text{if } C(h) = \set{(g,g')} 
     \tand 
     % \\
%     & \hspace{10mm} 
    h \text{ has a
      predecessor in } H\,.
    % \text{otherwise}, \\
    % & \hspace{10mm} \text{provided that } C(h) = \set{(g,g')}\,.
  \end{cases}
\end{align}
Let us remark that if $h \in \varsigma(g,g',h_{0})$, then $f(h) \in
\set{g,g'}$. If $gRh$ and $h$ has no predecessor, then $f(h) = g$.
Also, if $h'$ is the target of $\varsigma(g,g',h_{0})$ and $g'$ has a
successor, then $f(h') = g'$.
\begin{lemma}
  \label{LEMMA:SIM:COVER}
  If $(R,\varsigma)$ is a $\star$-weak simulation of $G$ by $H$ and
  $\rho : K \rTo H$ is an unravelling of $H$, then there exists a
  $\star$-weak simulation $(\tilde{R},\tilde{\varsigma})$ of $G$ by
  $K$.
\end{lemma}
Let us now recall that if $H$ is a tree with back edges, rooted at
$h_{0}$, of feedback $k$, then Cops has a \emph{canonical winning
  strategy} in the game $\Ent{H,k}$ from position $(h_{0},C,Cops)$.
%%
% Cops have the following winning strategy in the game $\Ent{H,k}$ (we
% shall refer to it as the ).
Every time a return is visited, a cop is dropped on such a return. If
a cop has to be replaced in order to occupy such a return, then the
cop which is closest to the root is chosen. 
\begin{remark}
  \label{remark:positionsremark}
  Let us remark that, by using the canonical strategy, (i) every path
  chosen by Thief in $H$ is a tree path, (ii) if the position in
  $\Ent{H,k}$ is of the form $(h,C,Thief)$, and $h'\neq h$ is in the
  subtree of $h$, then the unique tree path from $h$ to $h'$ does
  contain no cops, apart possibly for the vertex $h$.
% \begin{enumerate}
% \item every path chosen by Thief in $H$ is a tree path,
% \item if the position in $\Ent{H,k}$ is of the form $(h,C,Thief)$, and
%   $h'\neq h$ is in the subtree of $h$, then the unique path from $h$
%   to $h'$ does contain no cops, apart possibly for the vertex $h$.
% \end{enumerate}
  Finally, a vertex $h \in V_{H}$ determines a position
  $(h,C_{H}(h),Thief)$ in the game $\Ent{H,k}$ that has been reached
  from the initial position $(h_{0},\emptyset,Cops)$ and where Cops
  have been playing according to the canonical strategy.  $C_{H}(h)$ is
  determined as the set of returns $r$ of $H$ on the tree path from
  $h_{0}$ to $h$ such that the tree path from $r$ to $h$ contains at
  most $k$ returns.
\end{remark}

The following Theorem establishes the desired connection between
$\star$-weak simulations and entanglement.
\begin{theorem}
  \label{weaksim:entag:prop}
  If $(R,\varsigma)$ is a $\star$-weak simulation of $G$ by $H$, then
%   $$
%   \Ent{G} \leq \Ent{H} + 2\,.
%   $$
  $\Ent{G} \leq \Ent{H} + 2$.
\end{theorem}
\begin{proof}
  Let $k = \Ent{H}$.  We shall define first a strategy for Cops in the
  game $\ET{G,k+2}$.  In a second time, we shall prove that this
  strategy is a \emph{winning} strategy for Cops.

  Let us consider Thief's first move in $\ET{G,k + 2}$.  This move
  picks $g \in G$ leading to the position $(g,\emptyset,Cops)$ of
  $\ET{G,k +2}$. Cops answers by occupying the current position, i.e.
  he moves to $(g,\set{g},Thief)$.  After this move, Cops also chooses
  a tree with back edges of feedback $k$ to which $H$ unravel, $ \pi :
  \mathcal{T}(H) \rTo H $, such that the root $h_{0}$ of
  $\mathcal{T}(H)$ satisfies $gR\pi(h_{0})$. We can also suppose that
  $h_{0}$ is not a return, thus it has no predecessor. According to
  Lemma \ref{LEMMA:SIM:COVER} we can lift the $\star$-weak simulation
  $(R,\varsigma)$ to a $\star$-weak simulation
  $(\tilde{R},\tilde{\varsigma})$ of $G$ by $\mathcal{T}(H)$. In other
  words, we can suppose from now on that $H$ itself is a tree with
  back edges of feedback $k$ rooted at $h_{0}$ and, moreover, that $g
  R h_{0}$.

  From this point on, Cops uses a memory to choose how to place cops in
  the game $\ET{G,k+2}$. To each Thief's position $(g,C_{G},Thief)$
  in $\ET{G,k+2}$ we associate a data structure (the memory)
  consisting of a triple $M(g,C,Thief) = (p,c,h)$, where $c,h \in
  V_{H}$ and $p \in V_{H} \cup \set{\bot}$ (we assume that $\bot
  \not\in V_{H}$). Moreover $c \above h$ in the tree and, whenever $p
  \neq\bot$, $p \above c$ as well.

  Intuitively, we are matching the play in $\ET{G,k +2}$ with a play
  in $\Ent{H,k}$, started at the root $h_{0}$ and played by Cops
  according to the canonical strategy. Thus $c$ is the vertex of $H$
  currently occupied by Thief in the game $\Ent{H,k}$.%
  \footnote{More precisely we are associating to the position
    $(g,C_{G},Thief)$ of $\Ent{G,k+2}$ the position $(c,C_{H},Thief)$
    in $\Ent{H,k}$, where $C_{H}$ is determined as $C_{H} = C_{H}(c)$
    as in Remark \ref{remark:positionsremark}.}  Instead of recalling
  all the play (that is, the history of all the positions played so
  far), we need to record the last position played in $\Ent{H,k}$:
  this is $p$, which is undefined when the play begins.  Cops on $G$
  are positioned on the images of Cops on $H$ by the function $f$
  defined in \eqref{eq:deff}. Moreover, Cops eagerly occupies the last
  two vertices visited on $G$.  Thief's moves on $G$ are going to be
  simulated by sequences of Thief's moves on $H$, using the
  $\star$-weak simulation $(R,\varsigma)$. In order to make this
  possible, a simulation of the form
  $\varsigma(\tilde{g},g,\tilde{h})$ must be halted before its target
  $h$; the current position $c$ is such halt-point. This implies that
  the simulation of $g \rightarrow g'$ by $(R,\varsigma)$ and the
  sequence of moves in $H$ matching Thief's move on $G$ are sligthly
  out of phase.  To cope with that, Cops must guess in advance what
  might happen in the rest of the simulation and this is why he puts
  cops on the current and previous positions in $G$.  We also need to
  record $h$, the target of the previous simulation into the memory.
  
  The previous considerations are formalized by requiring the
  following conditions to hold. To make sense of them, let us say that
  $f(\set{p}) = f(p)$ if $p \in V_{H}$ and that $f(\set{p}) =
  \emptyset$ if $p = \bot$. In the last two conditions we require that
  $p \neq \bot$.
  \begin{align*}
    \bullet \;\;& C_{G}  = f(C_{H}(c)) \cup f(\set{p}) \cup \set{g}
    \,,
    \mbox{\hspace{4cm}}
    \tag{COPS}
    \label{cond:COPS}
    \\
    \bullet \;\;& f(c)  = g, \tand f(h') \in f(\set{p}) \cup
    \set{g}, \text{ whenever}\\
    \label{cond:tail}
    \tag{TAIL}
    & \hspace{10mm}%%
    \text{ $h'$ lies on the  tree path from $c$ to $h$}\,,
    \\%%\intertext{If $p \neq \bot$, then}
    \bullet \;\;& f(p) \rightarrow g\,, f(p)R\tilde{h} \text{ for some }
    \tilde{h}\in V_{H},
    c \in \varsigma(f(p),g,\tilde{h}),
    \\
    \tag{HEAD}
    \label{cond:head}
    & \hspace{10mm} \tand h \text{ is the target of
    }\varsigma(f(p),g,\tilde{h})\,, \\
    \bullet \;\;& \text{on the tree path from $p$ to $c$},
    \\
    \tag{HALT}
    \label{cond:cutpoint}
    &
    \hspace{10mm}
    \text{ $c$ is the only vertex} \tst f(c) = g \,.
  \end{align*}

  Since $h_{0}$ has no predecessors, then $gRh_{0}$ implies $f(h_{0})
  = g$.  Thus, at the beginning, the memory is set to
  $(\bot,h_{0},h_{0})$ and conditions \eqref{cond:COPS} and
  \eqref{cond:tail} hold.

  Consider now a Thief's move of the form $(g,C_{G},Thief) \rightarrow
  (g',C_{G},Cops)$, where $g' \not\in C_{G}$. If $g'$ has no
  successor, then Cops simply skips, thus reaching a winning position.
  Let us assume that $g'$ has a successor, and  
  %% Let 
  write $\varsigma(g,g',h) = h h_{1}\ldots h_{n}$, $n \geq 1$;
  observe that $f(h_{n}) = g'$.  If for some $i = 1,\ldots ,n$ $h_{i}$
  is not in the subtree of $c$, then the strategy halts, Cops abandons
  the game and looses.  Otherwise, all the path $\pi = c\ldots h h_{1}
  \ldots h_{n}$ lies in the subtree of $c$. By eliminating cycles from
  $\pi$, we obtain a simple path $\sigma$, of source $c$ and target
  $h_{n}$, which entirely lies in the subtree of $c$.  By Lemma
  \ref{lemma:simplepathtwbe}, $\sigma$ is the tree path from $c$ to
  $h_{n}$. An explicit description of $\sigma$ is as follows: we can
  write $\sigma$ as the compose $\sigma_{0} \star \sigma_{1}$, where
  the target of $\sigma_{0}$ and source of $\sigma_{1}$ is the vertex
  of $\varsigma(g,g',h)$ which is closest to the root $h_{0}$;
  moreover $\sigma_{0}$ is a prefix of the tree path from $c$ to $h$,
  and $\sigma_{1}$ is a postfix of the path $\varsigma(g,g',h)$.

  We cut $\sigma$ as follows: we let $c'$ be the first vertex on this
  path such that $f(c') = g'$.
  Thief's move $g \rightarrow g'$ on $G$ is therefore simulated by
  Thief's moves from $c$ to $c'$ on $H$. This is possible
  since every vertex lies in the subtree of $c$ and thus it has not
  yet been explored. Cops consequently occupies the returns on this
  path, thus modifying $C_{H}$ to $C'_{H} = C_{H}(c') = (C_{H}
  \setminus X) \uplus Y$, where $Y$ is a set of at most $k$ vertexes
  containing the last returns visited on the path from $c$ to $c'$.

  After the simulation on $H$, Cops moves to $(g',C_{G}',Thief)$ in
  $\ET{G,k+2}$, where $C'_{G} = f(C'_{H}) \cup \set{g,g'}$.  Let us
  verify that this is an allowed move according to the rules of the
  game.  We remark that
%   \begin{align*}
%     C_{H}' & = (C_{H} \setminus X) \uplus Y
%   \end{align*}
  $f(Y) \subseteq f(\set{p}) \cup \set{g,g'}$ and therefore
  \begin{align*}
    C'_{G} & = f(C_{H}\setminus X) \cup f(Y) \cup \set{g,g'} \\
    & = (f(C_{H}\setminus X) \cup (f(Y)\setminus \set{g'}) \cup
    \set{g}) \cup \set{g'}\\
    & = A \cup \set{g'}\,, %%
    % \intertext{where} A & = f(C_{H}\setminus X) \cup (f(Y)\setminus
%     \set{g'}) \cup \set{g} %
%     \subseteq f(C_{H}) \cup f(\set{p}) \cup \set{g} = C_{G}\,.
  \end{align*}
  where $A = f(C_{H}\setminus X) \cup (f(Y)\setminus \set{g'}) \cup
  \set{g} \subseteq f(C_{H}) \cup f(\set{p}) \cup \set{g} = C_{G}$.
  After the simulation Cops also updates the memory to
  $M(g',C_{G}',Thief) = (c,c',h_{n})$. Since $f(c) = g$, then
  condition \eqref{cond:COPS} clearly holds. Also, $f(c) = g
  \rightarrow g'$, $g R h$ and $h_{n}$ is the target of
  $\varsigma(f(c),g',h)$. We have also that $c' \in \sigma_{1}$ and
  hence $c' \in \varsigma(f(c),g',h)$, since otherwise $c' \in
  \sigma_{0}$ and $f(c') \in \set{f(p),g}$, contradicting $f(c') = g'$
  and the condition on the girth of $G$.  Thus condition
  \eqref{cond:head} holds as well.  Also, condition
  \eqref{cond:cutpoint} holds, since by construction $c'$ is the first
  vertex on the tree path from $c$ to $h$ such that $f(c') = g'$. Let
  us verify that condition \eqref{cond:tail} holds: by construction
  $f(c') = g'$, and the path from $c'$ to $h_{n}$ is a postfix of
  $\varsigma(g,g',h)$, and hence $f(h') \in \set{g,g'}$ if $h'$ lies
  on this tree path.

%   Therefore let us show that after having simplified the path 
%   $(c\ldots \source\pi \ldots \target\pi)$, if $i < j$ and $f(h_{i}) =
%   g'$, then $f(h_{j}) \in \set{g,g'}$.
  
%   Let $r$ be the top of simple path $\pi$, so that $\pi = \eta \ast
%   \tau(\pi)$ with $r = \source(\tau(\pi)) = \target (\eta)$.  Then $r$
%   lies on the tree-path from $c$ to $h$. Therefore the desired simple
%   path is of the form $\pi_{c,r} \ast \tau(\pi)$, where $\pi_{c,r}$ is
%   a prefix of $\pi_{c,h}$, and therefore for $h_{i} \in \pi_{c,r}$ we
%   have $f(h_{i}) \in \set{f(p),g}$. Therefore, since $g' \not\in
%   \set{f(p),g}$, if $h_{i}$ is such that $f(h_{i}) = g'$, then $h_{i}
%   \in \tau(\pi)$. If $i < j$, then also $h_{j} \in \tau(\pi)$, so that
%   $f(h_{j}) \in \set{g,g'}$.

%   The memory is obviously updated to $M(g',C_{G}',T) =
%   (\set{c},c',h')$. Let us also say that if any of the conditions
%   under which the memory is subject fails, then Cops abandons and loose
%   the game.
  
  Let us now prove that the strategy is winning. If Cops never
  abandons, then an infinite play in $\ET{G,k+2}$ would give rise to
  an infinite play in $\Ent{H,k}$, a contradiction.  Thus, let us
  prove that Cops will never abandon. To this goal we need to argue
  that when Thief plays the move $g \rightarrow g'$ on $G$, then the
  simulation $\varsigma(g,g',h) = h h_{1}\ldots h_{n}$ lies in the
  subtree of $c$.  If this is not the case, let $i$ be the first index
  such that $h_{i}$ is not in the subtree of $c$. Therefore $h_{i}$ is
  a return and, by the assumptions on $H$ and the on canonical
  strategy, $h_{i} \in C_{H}(c)$.  Since $h_{i} \in
  \varsigma(g,g',h)$, $f(h_{i}) \in \set{g,g'}$. Observe, however that
  we cannot have $f(h_{i}) = g'$, otherwise $g' \in f(C_{H}(c))
  \subseteq C_{G}$.  We deduce that $f(h_{i}) = g$ and that $g \in
  f(C_{H}) \subseteq C_{G}$.

  Since $C_{G} \neq \bot$, then
  $(g,C_{G},Thief)$ is not the initial position of the play, so that,
  if $M(g,C_{G},Thief) = (p,c,h)$, then $p \neq \bot$.
  Let us now consider the last two moves of the play before
  reaching position 
  $(g,C_{G},Thief)$. These are of the form
%   \begin{align*}
%     (f(p),\tilde{C}_{G},Thief)
%     & \rightarrow 
%     (g,\tilde{C}_{G},Cops)
%     \rightarrow 
%     (g,C_{G},Thief)\,,
%   \end{align*}
  $(f(p),\tilde{C}_{G},Thief) \rightarrow (g,\tilde{C}_{G},Cops)
  \rightarrow (g,C_{G},Thief)$, %%
  and have been played according to this strategy. Since $g \not\in
  \tilde{C}_G$, it follows that the Cop on $h_{i}$ has been dropped on
  $H$ during the previous round of the strategy, simulating the move
  $f(p) \rightarrow g$ on $G$ by the tree path from $p$ to $c$.  This
  is however in contradiction with condition \eqref{cond:cutpoint},
  stating that $c$ is the only vertex $h$ on the tree path from $p$ to
  $c$ such that $f(h) = c$.
\end{proof}
%% \myqed
% leading from  the
%  that is,
%   it is a return that lies on the path from $p$ to $c$. However, this
%   is a contradiction, since by assumption $c$ is the first vertex on
%   this path such that $f(c) = g$, and $f(h_{i}) = g$ is above this
%   vertex.
 %  Thus we are only left to prove that the invariants defining
%   conditions on $M(g',C_{G}',T) = (\set{c},c',h')$ are respected. A
%   part from the last one, all the invariants holds by construction.

%   Therefore let us show that after having simplified the path 
%   $(c\ldots \source\pi \ldots \target\pi)$, if $i < j$ and $f(h_{i}) =
%   g'$, then $f(h_{j}) \in \set{g,g'}$.
  
%   Let $r$ be the top of simple path $\pi$, so that $\pi = \eta \ast
%   \tau(\pi)$ with $r = \source(\tau(\pi)) = \target (\eta)$.  Then $r$
%   lies on the tree-path from $c$ to $h$. Therefore the desired simple
%   path is of the form $\pi_{c,r} \ast \tau(\pi)$, where $\pi_{c,r}$ is
%   a prefix of $\pi_{c,h}$, and therefore for $h_{i} \in \pi_{c,r}$ we
%   have $f(h_{i}) \in \set{f(p),g}$. Therefore, since $g' \not\in
%   \set{f(p),g}$, if $h_{i}$ is such that $f(h_{i}) = g'$, then $h_{i}
%   \in \tau(\pi)$. If $i < j$, then also $h_{j} \in \tau(\pi)$, so that
%   $f(h_{j}) \in \set{g,g'}$.
%  \myqed
% \end{proof}

%%% Local Variables: 
%%% mode: latex
%%% TeX-master:"0"
%%% End: 

\section{Strongly Synchronizing Games} 
\label{sec:stronggames}

In this section we define  \emph{strongly
  synchronizing} games,
% a kind of asymmetric objects which are hard
% w.r.t the variable hierarchy.
%These games are 
a generalization of synchronizing games introduced in \cite{TAC02}.
% as a fundamental tool to proof that the alternation depth of games
% $\mu$-calculus is strict.
We shall show that, for every game $H$ equivalent to a strongly
synchronizing game $G$, there is a $\star$-weak simulation of $G$ by
$H$.\footnote{In the sequel, we shall not distinguish between a game
  and its underlying graph.}

Let us say that $G \in {\cal G}$ is \emph{bipartite} if $M^G \subseteq
Pos^G_{E} \times Pos^G_{A,D} \; \cup \; Pos^G_{A} \times Pos^G_{E,D}$.
\begin{definition} \label{StrongSynch} A game $G$ is \emph{strongly
    synchronizing} iff its is bipartite, it has girth strictly greater
  than $4$ and, for every pair of positions $g,k$, the following
  conditions hold:
 \begin{enumerate}
 \item if $(G,g) \sim (G,k)$  then $g = k$.
 \item if $(G,g) \leq (G,k)$ and $(G,k) \not\leq (G,g)$, then $k \in
   Pos_{E}^G \tand (k,g) \in M^G$, or $g \in Pos_{A}^G \tand (g,k) \in
   M^G$.
%  implies
%    \begin{align*}
%      k   & \in Pos_{E}^G \tand (k,g) \in M^G\,, 
%      & \tor 
%      \;\;\;\;
%       g & \in Pos_{A}^G  \tand (g,k) \in M^G\,.
%    \end{align*}
 \end{enumerate}
\end{definition}
A consequence of the previous definition is that \emph{the only
  winning strategy for Mediator in the game $\homl G,G \homr$ is the
  copycat strategy}. Thus strongly synchronizing games are
synchronizing as defined in \cite{TAC02}.  We list next some useful
properties of strongly synchronizing games.
\begin{lemma} \label{Lemma:Intersec} 
  %\label{intersection1:lemma} 
  %\label{intersection2:lemma}
  Let $G$ be a strongly synchronizing  and
  let $(g,g'),(\tilde{g}, \tilde{g}') \in M^G$ be distinct.
  \begin{enumerate}
  \item   If $(G,g) \sim \hat{x}$ then $g \in
    Pos^{G}_{D}$ and $\lambda(g) = x$.
  \item If $g,\tilde{g} \in Pos_{E}^G$ and,
    for some game $H$ and $h \in Pos^{H}$, we have
    \begin{align*}
      (G,g') & \leq (H,h) \leq (G,g) 
      %&%
      \tand\\ %&
      & \myhspace{10mm}
      (G,\tilde{g}') %& 
      \leq (H,h) \leq (G,\tilde{g})\,,
    \end{align*}   
    then $g=\tilde{g}$ or $g'=\tilde{g}'$, and
    $|\set{g,g',\tilde{g},\tilde{g}'}|=3$.
  \item If $g \in Pos_{E}^G$ and $\tilde{g} \in Pos_{A}^G$ and, for some
    $H$ and $h \in Pos^H$, we have
    \begin{align*}
      (G,g') & \leq (H,h) \leq (G,g) 
      %&
      %% 
      \tand 
      \\%&
      & \myhspace{10mm}
      (G,\tilde{g}) %& 
      \leq (H,h) \leq (G,\tilde{g}')\,,
    \end{align*}
    then $g=\tilde{g}'$ or $g'=\tilde{g}$, and
    $|\set{g,g',\tilde{g},\tilde{g}'}|=3$.
  \end{enumerate}
\end{lemma}

% \begin{lemma}  Let $(g,g'),(\tilde{g}, \tilde{g}') \in M^G$  such that
%   \begin{itemize}
%   \item  $g \in Pos_{E}^G$ and $\tilde{g} = Pos_{A}^G$,
%   \item for some $H$ and $h \in Pos^H$, we have
%     \begin{align*}
%       (G,g') & \leq (H,h) \leq (G,g) 
%       &
%       %% 
%       \tand &
%       %%
%       &
%       (G,\tilde{g}) & \leq (H,h) \leq (G,\tilde{g}')\,.
%     \end{align*}
%   \end{itemize}
%   Then $g=\tilde{g}'$ or $g'=\tilde{g}$, i.e  $|\set{g,g',\tilde{g},\tilde{g}'}|=3$
% \end{lemma}

\noindent
We are ready to state the main result of this section.
\begin{proposition} % 
  \label{weaksim:game:prop} 
  Let $G$ be a strongly
  synchronizing game, and let $H \in \mathcal{G}$ be such that $G \leq
  H \leq G$, then there is a $\star$-weak simulation of $G$ by $H$.
\end{proposition}
\begin{proof}
  Let $S,S'$ be two winning strategies for Mediator in $\langle G,H
  \rangle$ and $\langle H, G\rangle$, respectively.  Let $T = S || S'$
  be the composal strategy in $\langle G,H,G \rangle$. Define
  \begin{align*}
    g R h & \tiff % g \in Pos_{P}^G, h \in Pos_{P}^H, \textrm{ where } P\in\set{E,A,D} \tand \\
%      & \hspace{15mm}
    (g,h,g)
    \text{ is a position of } T\\
    %\text{ of the form } (g,h,g)\, 
    & \hspace{5mm}\tand g,h \text{ belong to the same player}.
  \end{align*}
  We consider first $R$ and prove that it is functional and
  surjective.  If $ g_i R h, i=1,2$ then $(g_1,h,g_1)$ and
  $(g_2,h,g_2)$ are positions of $T$, hence $(G,g_1) \le (H,h) \le
  (G,g_1)$ and $(G,g_2) \le (H,h) \le (G,g_2)$, consequently $(G,g_1)
  \sim (G,g_2)$ implies $g_1=g_2$, by definition \ref{StrongSynch}.
  For surjectivity, we can assume that (a) all the positions of $G$
  are reachable from the initial position $p_{\star}^G$, (b)
  $p_{\star}^G$ and $p_{\star}^H$ belong to the same player (by
  possibly adding to $H$ a new initial position leading to the old
  one).  Since $T_{\setminus H }$ is the copycat strategy, given $g
  \in Pos^{G}_{E,A,D}$, from the initial position
  $(p_{\star}^G,p^{H}_{\star},p_{\star}^G)$ of $ \homl G,H,G \homr$,
  the Opponents have the ability to reach a position of the form
  $(g,h,g)$.
  % Therefore, since $G \le H \le G$, for each $g \in
  % Pos^{G}$ there is some $h \in Pos^{H}$ such that $(G,g)\sim (H,h)$.
  The explicit construction of the function $\varsigma$ will show that
  $h$ can be chosen to belong to the same player as $g$.

%   We shall prove that $(R,\varsigma)$ is a $\star$-weak simulation, we
%   shall make precise the function $\varsigma$ by the next.

  We construct now the function $\varsigma$ so that $(R,\varsigma)$ is
  a weak simulation.  If $g R h$ and $(g,g') \in M^G$, then we
  construct $\pi= h,\dots, h'$ such that $g' R h'$. Since $G$ is
  bipartite, then $h \neq h'$ and $\pi$ is nonempty.  We let
  $\varsigma (g,g',h)$ be a reduction of $\pi$ to a nonempty simple
  path.

  We assume $(g,h) \in (Pos_{E}^G,Pos_{E}^H)$, the case
  $(g,h) \in (Pos_{A}^G,Pos_{A}^H)$ is dual.  From position $(g,h,g)$
  it is Opponent's turn to move on the left, they choose a move
  $(g,g') \in M^G$. Since $G$ is bipartite, we have either $g' \in
  Pos_{D}^G$ or $g' \in Pos_{A}^G$.
  \begin{proofbycases}
    %\vskip -2pt
    \begin{caseinproof}
      If $g' \in Pos_{D}^G$ then the strategy $T$ suggests playing a
      finite path on $H$, $(g',h,g) \to^* (g',h^*,g)$, possibly of
      zero length, and then it will suggest to play on the external
      right board. An infinite path played only on $H$ cannot arise,
      since $T$ is a winning strategy and such an infinite path is not
      a win for Mediator. Since $T_{\setminus H}$ is the copycat
      strategy, $T$ suggests the only move $(g',h^*,g) \to
      (g',h^*,g')$.  From this position $T$ suggests playing a path on
      $H$ leading to a final draw position $h_f \in Pos_{D}^H$ as
      follows $(g',h^*,g') \to^* (g',h_f,g')$, such that
      $\lambda^G(g')=\lambda^H(h_f)$, therefore $g' R h_f$.
 \end{caseinproof}
 \begin{caseinproof} 
   If $g' \in Pos_{A}^G$ then from position $(g',h,g)$ it is
   Mediator's turn to move.  We claim that $T$ will suggest playing a
   nonempty finite path $(g',h,g) \to^{+} (g',h',g)$ on the central
   board $H$, where $h' \in Pos_{A}^H$, and then suggests the move
   $(g',h',g) \rightarrow (g',h',g')$.  Let $\tilde{h} \in
   Pos^{H}_{A,E,D}$ be such that the position $(g',\tilde{h},g)$ has
   been reached from $(g',h,g)$, through a (possibly empty) sequence
   of central moves, by playing with $T$.
   Then $T$ cannot suggest a move on the left board
   $(g',\tilde{h},g) \to (g'',\tilde{h},g)$, since $T_{\setminus H}$
   is the copycat strategy. Also, if $\tilde{h} \in Pos^{H}_{E}$, $T$
   cannot suggest a move on the right board $(g',\tilde{h},g) \to
   (g',\tilde{h},\tilde{g})$. The reason is that $T = S \comps S'$, and
   the position $(\tilde{h},g)$ of $\homl H,G\homr$ does not allow a
   Mediator's move on the right board. Thus a sequence of central
   moves on $H$ is suggested by $T$ and, as mentioned above, this
   sequence cannot be infinite. We claim that its endpoint $h'
   \in Pos^{H}_{A}$.  We already argued that $h' \not\in
   Pos^{H}_{E}$, let us argue that $h' \not\in Pos^{H}_{D}$. If
   this were the case, then strategy $T$ suggests the only move
   $(g',h',g) \to (g',h_n,g')$, hence $(G,g') \sim (H,h')$. By Lemma
   \ref{Lemma:Intersec}.1, we get $g' \in Pos_{D}^G$, contradicting $g' \in
   Pos_{A}^G$.
%%
%    Indeed $h_n \in Pos_{A}^H$,
%    because if $h_n \in Pos_{E}^H$, then the strategy $T$ can not
%    suggest a move on the left board $(g',h_n,g) \to (g'',h_n,g)$ --
%    since $T_{\setminus H}$ is the copycat strategy -- moreover the
%    right strategy $S$ forces Opponents to move on $H$ from position
%    $(h_n,g)$ of the game $\langle H,G \rangle$.  It can not also be
%    the case $h_n \in Pos_{D}^H$, because if it is the case the
%    strategy $T$ suggests the only move $(g',h_n,g) \to (g',h_n,g')$,
%    hence $(G,g') \sim (H,h_n)$. By Lemma \ref{SynchFinal}, we get $g'
%    \in Pos_{D}^G$, that contradicts the assumption $g' \in Pos_{A}^G$.
%    %%
%    From position $(g',h_n,g)$ it is Mediator's turn to move, $T$ can
%    not suggest to play on the left board, otherwise the strategy
%    $T_{\setminus H}$ is not the copycat strategy, so it suggests a
%    move on the right board. Since $T_{\setminus H}$ is the copycat,
%    the only possible move is $(g',h',g) \to (g',h',g')$ and hence $g'
%    R h'$.
\end{caseinproof}
\end{proofbycases}
This proves that $(R,\varsigma)$ is a weak simulation.  We prove next
that $(R,\varsigma)$ has the $\star$-property, thus assume that $h^{*}
\in
\varsigma(g,g',h_{0}),\varsigma(\tilde{g},\tilde{g}',\tilde{h}_{0})$.
Let us suppose first that $g,\tilde{g} \in Pos_{E}^H$.  By looking at
the construction of these paths, we observe that the two sequences of
moves
\begin{align*}
  (g,h_0,g) & \to (g',h_0,g) \to^* (g',h^*,g) 
  % \\
%   &  \hspace{5mm}
  \to^* (g',h_n,g) \to (g',h_n,g')  \,,\\
  (\tilde{g},\tilde{h}_0,\tilde{g}) & \to
  (\tilde{g}',\tilde{h}_0,\tilde{g}) \to^* (\tilde{g}',h^*,\tilde{g})
  % \\
  % & \hspace{5mm}
  \to^* (\tilde{g}',\tilde{h}_m,\tilde{g}) \to
  (\tilde{g}',\tilde{h}_m,\tilde{g}')\,,
\end{align*} 
%\begin{align*}
%   & (g,h_0,g) \to (g',h_0,g) \to^* (g',h^*,g) \to^* (g',h_n,g) \to (g',h_n,g')  \\
%   & (\tilde{g},\tilde{h}_0,\tilde{g}) \to
%   (\tilde{g}',\tilde{h}_0,\tilde{g}) \to^* (\tilde{g}',h^*,\tilde{g})
%   \to^* (\tilde{g}',\tilde{h}_m,\tilde{g}) \to
%   (\tilde{g}',\tilde{h}_m,\tilde{g}')
% \end{align*} 
may be played in the game $\homl G,H,G \homr$, according to the winning
strategy $T = S \comps S'$.  We have therefore that $(G,g') \le (H,h^*)
\le (G,g)$ and $(G,\tilde{g}') \le (H,h^*) \le (G,\tilde{g})$.%
\footnote{%
  Similar inequalites may be derived even if $h^{*} \in Pos^{H}_{D}$.
  In this case the moves in the central board may be 
  interleaved with the move on the right board.} %
Consequently $|\set{g,g',\tilde{g},\tilde{g}'}|=3$, by Lemma
\ref{Lemma:Intersec}.2. If $g \in Pos_{E}^G$ and $\tilde{g} \in
Pos_{A}^G$, a similar argument shows that the positions $(g',h^*,g)$
and $(\tilde{g},h^*,\tilde{g}')$ may be reached with $T$ and hence
$(G,g') \le (H,h^*) \le (G,g)$ and $(G,\tilde{g}) \le (H,h^*) \le
(G,\tilde{g}')$.  Lemma \ref{Lemma:Intersec}.3 implies then
$|\set{g,g',\tilde{g},\tilde{g}'}|=3$.  Finally, the cases
$(g,\tilde{g}) \linebreak \in \set{(Pos_{A}^G,Pos_{A}^G),(Pos_{A}^G,Pos_{E}^G) }$
are handled by duality.
%%
%\noindent\hfill
This completes the proof of Proposition
\ref{weaksim:game:prop}.  \myqed
\end{proof}

\newpage
\section{Construction of Strongly Synchronizing Games }
\label{sec:ourgames}
In this section we complete the hierarchy theorem by constructing, for
$n \geq 1$, strongly synchronizing games $G_n$ such that $\Ent{G_{n}}
= n$. This games mimic the $n$-cliques already used in
\cite{BerwangerGraLen02} to prove that the variable hierarchy for the
modal $\mu$-calculus is infinite. 
% We begin by drawing  the game
% $G_{2}$:
The game $G_{2}$ appears in Figure 1.
\begin{figure*}[t]
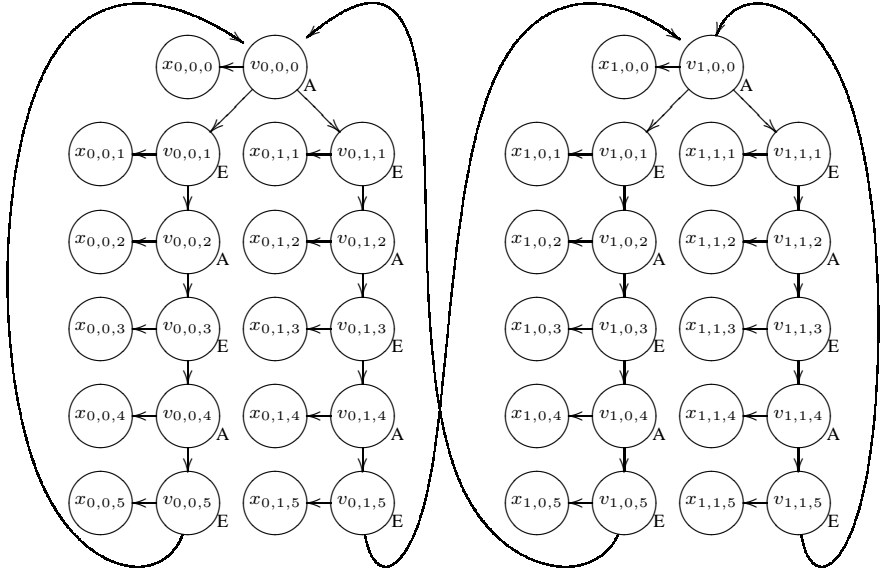

  \centering
$$
\let\objectstyle=\scriptstyle
\myxygraph[3.3em]{
  []!p{v_{0,0,0}}(:[l]!o{x_{0,0,0}})
  "P"="S0"
  (
  :[dl]!g{0,0},
  :[dr]!g{0,1}
  )
  [r(5)]!p{v_{1,0,0}}
  "P"="S1"
  (:[l]!o{x_{1,0,0}})
  (
  :[dl]!g{1,0},
  :[dr]!g{1,1}
  )
  "E_{0,0}"([d]="1"[l(2.2)]="2"[u(7)]="3"[r(2)]="4")
  :@`{"1","2","3","4"}"S0"
  "E_{0,1}"([d]="1"[r]="2"[u(7)]="3"[rr]="4")
  :@`{"1","2","3","4"}"S1"
  "E_{1,0}"
  ([d]="1"[l(2.6)]="2"[u(7)r(0.6)]="3"[l(1)u(0.01)]="4")
  :@`{"1","2","3"
    ,"4"%%}
  }"S0"
  "E_{1,1}"([d]="1"[r]="2"[u(7)]="3"[ll]="4")
  :@`{"1","2","3","4"}"S1"
}
$$
\caption{The game $G_{2}$}
\end{figure*}

The general definition of the game $G_{n}$ is as follows.  Let $[n]$
denote the set $\set{0,\ldots ,n-1}$ and let $I_{n} = \set{ (i,j,k)
  \in [n] \times [n] \times [6]\mid k = 0 \timplies j =0 }$.
%
%  \begin{align*}
%   X & = \set{x_{i} \mid i \geq 0 } &
%   \widetilde{X} & = \set{x_{i,j,k}\mid i,j\geq 0,k=1,\ldots, 5} 
% \end{align*}
% be two infinite sets of variables.
%
We define 
\begin{align*}
  Pos_{A}^{G_n} = & \;\; \set{v_{i,j,k} \mid (i,j,k) \in I_{n} 
    \tand k \mymod 2
    = 0} \,, \\
  Pos_{E}^{G_n}= & \;\;\set{v_{i,j,k} \mid (i,j,k) \in I_{n} \tand k
    \mymod 2
    = 1} \,, \\
  Pos_{D}^{G_n}= & \;\; \set{w_{i,j,k} \mid (i,j,k) \in I_{n}}\,.
\end{align*}
Let $X = \set{x_{i,j,k}\mid i,j\geq 0,k \in [n]}$ be a countable set
of variables, the labelling of draw positions, $\lambda^{G_n} :
Pos_D^{G_n} \rTo X$, sends $w_{i,j,k}$ to $x_{i,j,k}$. % as
% follows.
% \begin{align*}
%   & \lambda^{G_n} (w_{i})= x_{i} &  \tand& &\lambda^{G_n}(w_{i,j,k}) = x_{i,j,k}\,.
% \end{align*}
The moves $M^{G_n}$ either lie on some cycle:
\begin{align*}
  v_{i,0,0} & \rightarrow v_{i,j,1}, 
  &v_{i,j,k} \rightarrow v_{i,j,k + 1}, & \;\;k = 1,\ldots ,4, \\
  v_{i,j,5} & \rightarrow v_{j,0,0}\,, 
%   \cup \set{(v_{i,j,k},w_{i,j,k}),(v_{i,j,k},v_{i,j,k + 1}) \mid i,j
%     \in [n], \, k = 1,\ldots ,4 } \\ 
%   \intertext{or lead to draw positions:}
%   & \set{(v_{i},w_{i}),(v_{i,j,k},w_{i,j,k}) \mid i,j
%     \in [n], \, k = 1,\ldots ,5}\,.
\end{align*}
or lead to draw positions: $v_{i,j,k} \rightarrow w_{i,j,k}$.
% \begin{align*}
%   M^{G_n}  = & \myhide{\cup}\set{(v_{i},w_{i}),(v_{i},v_{i,j,1}) \mid i \in
%     [n]} \\
%   & \cup \set{(v_{i,j,k},w_{i,j,k}),(v_{i,j,k},v_{i,j,k + 1}) \mid i,j
%     \in [n], \, k = 1,\ldots ,4 } \\ 
%   & \cup \set{(v_{i,j,5},w_{i,j,5}),(v_{i,j,5},v_{j}) \mid i,j
%     \in [n]}\,.
% \end{align*}
Finally, the priority function $\rho^{G_n}$ assigns a constant odd
priority to all positions. We state next the main facts about the
games $G_{n}$:
\begin{proposition}
  The games $G_n$ are strongly synchronizing and $\Ent{G_{n}} = n$.
\end{proposition}
The proof of the statement is omitted for lack of space.  We are now
ready to state the main achievement of this paper.
\begin{theorem} %
  \label{main:theorem}
  For $n \geq 3$, the inclusions $\mathcal{L}_{n-3} \subseteq
  \mathcal{L}_{n}$ are strict. Therefore the variable hierarchy for
  the games $\mu$-calculus is infinite.
\end{theorem}
By the previous Proposition the game $G_{n} \in \mathcal{L}_{n}$.
Also, since $G_{n}$ is strongly synchronizing, if $H \sim G_{n}$, then
there exists a $\star$-weak simulation of $G_{n}$ by $H$. It follows
by Theorem \ref{weaksim:entag:prop} that $n -2 \leq \Ent{H}$.
Therefore $G_{n} \not \in \mathcal{L}_{n-3}$.

%%% Local Variables: 
%%% mode: latex
%%% TeX-master: "0"
%%% End: 

%\bibliographystyle{splncs}
%\bibliographystyle{plain}
\bibliographystyle{latex8}
\bibliography{biblio}

\begin{thebibliography}{10}\setlength{\itemsep}{-1ex}\small

\bibitem{Aabram94}
S.~Abramsky and R.~Jagadeesan.
\newblock Games and full completeness for multiplicative linear logic.
\newblock {\em J. Symb. Logic}, 59(2):543--574, 1994.

\bibitem{AN}
A.~Arnold and D.~Niwi{\'n}ski.
\newblock {\em Rudiments of {$\mu$}-calculus}, volume 146 of {\em Studies in
  Logic and the Foundations of Mathematics}.
\newblock North-Holland, 2001.

\bibitem{FOSSACS03}
A.~Arnold and L.~Santocanale.
\newblock Ambiguous classes in the games {$\mu$}-calculus hierarchy.
\newblock In {\em FOSSACS 2003}, volume 2620 of {\em Lect. Not. Comp. Sci.},
  pages 70--86. Springer, 2003.

\bibitem{BelkSanto07}
W.~Belkhir and L.~Santocanale.
\newblock Undirected graphs of entanglement 2.
\newblock In {\em FSTTCS 2007}, volume 4855 of {\em Lect. Not. Comp. Sci.},
  pages 508--519. Springer, 2007.

\bibitem{BerwangerThesis}
D.~Berwanger.
\newblock {\em Games and Logical Expressiveness}.
\newblock PhD thesis, RWTH Aachen, 2005.

\bibitem{berwanger}
D.~Berwanger and E.~Gr{\"a}del.
\newblock Entanglement---a measure for the complexity of directed graphs with
  applications to logic and games.
\newblock In {\em LPAR 2005}, volume 3452 of {\em Lect. Not. Comp. Sci.}, pages
  209--223. Springer, 2005.

\bibitem{BerwangerGraLen02}
D.~Berwanger, E.~Gr{\"a}del, and G.~Lenzi.
\newblock On the variable hierarchy of the modal mu-calculus.
\newblock In {\em CSL 2002}, volume 2471 of {\em Lect. Not. Comp. Sci.}, pages
  352--366. Springer, 2002.

\bibitem{BerwangerLen05}
D.~Berwanger and G.~Lenzi.
\newblock The variable hierarchy of the $\mu$-calculus is strict.
\newblock In {\em STACS 2005}, volume 3404 of {\em Lect. Not. Comp. Sci.},
  pages 97--109. Springer, 2005.

\bibitem{blass72}
A.~Blass.
\newblock Degrees of indeterminacy of games.
\newblock {\em Fund. Math.}, 77(2):151--166, 1972.

\bibitem{Blass92}
A.~Blass.
\newblock A game semantics for linear logic.
\newblock {\em Ann. Pure Appl. Logic}, 56(1-3):183--220, 1992.

\bibitem{bloomesik}
S.~L. Bloom and Z.~{\'E}sik.
\newblock {\em Iteration theories}.
\newblock Springer, 1993.

\bibitem{Bradfield98}
J.~C. Bradfield.
\newblock The modal $\mu$-calculus alternation hierarchy is strict.
\newblock {\em Theor. Comput. Sci.}, 195(2):133--153, 1998.

\bibitem{cockettseely}
J.~R.~B. Cockett and R.~A.~G. Seely.
\newblock Finite sum-product logic.
\newblock {\em Theory Appl. Categ.}, 8:63--99 (electronic), 2001.

\bibitem{freese}
R.~Freese, J.~Je{\v{z}}ek, and J.~B. Nation.
\newblock {\em Free lattices}, volume~42 of {\em Mathematical Surveys and
  Monographs}.
\newblock American Mathematical Society, 1995.

\bibitem{joy77}
A.~Joyal.
\newblock Remarques sur la th\'{e}orie des jeux \`{a} deux personnes.
\newblock {\em La Gazette des Sciences Math\'{e}matiques du Qu\'{e}bec}, 1(4),
  March 1977.

\bibitem{Joyal95}
A.~Joyal.
\newblock Free bicomplete categories.
\newblock {\em C. R. Math. Rep. Acad. Sci. Canada}, 17(5):219--224, 1995.

\bibitem{Joyal97}
A.~Joyal.
\newblock Free lattices, communication and money games.
\newblock In {\em Logic and scientific methods (Florence, 1995)}, volume 259 of
  {\em Synthese Lib.}, pages 29--68. Kluwer Acad. Publ., 1997.

\bibitem{kozen}
D.~Kozen.
\newblock Results on the propositional {$\mu $}-calculus.
\newblock {\em Theoret. Comput. Sci.}, 27(3):333--354, 1983.

\bibitem{Nerode92}
A.~Nerode, A.~Yakhnis, and V.~Yakhnis.
\newblock Concurrent programs as strategies in games.
\newblock In Y.~N. Moschovakis, editor, {\em Logic from Computer Science:
  Proc.\ of a Workshop}, pages 405--479. Springer, 1992.

\bibitem{Parikh}
M.~Pauly and R.~Parikh.
\newblock Game logic---an overview.
\newblock {\em Studia Logica}, 75(2):165--182, 2003.
\newblock Game logic and game algebra (Helsinki, 2001).

\bibitem{TAC02}
L.~Santocanale.
\newblock The alternation hierarchy for the theory of $\mu$-lattices.
\newblock {\em Theory and Applications of Categories}, 9:166--197, Jan. 2002.

\bibitem{Luigi}
L.~Santocanale.
\newblock A calculus of circular proofs and its categorical semantics.
\newblock In {\em FOSSACS 2002}, pages 357--371, 2002.

\bibitem{freemulat}
L.~Santocanale.
\newblock Free $\mu$-lattices.
\newblock {\em Journal of Pure and Applied Algebra}, 168(2-3):227--264, Mar.
  2002.

\bibitem{TCS333}
L.~Santocanale and A.~Arnold.
\newblock Ambiguous classes in {$\mu$}-calculi hierarchies.
\newblock {\em Theoret. Comput. Sci.}, 333(1-2):265--296, Mar. 2005.

\bibitem{tarski}
A.~Tarski.
\newblock A lattice-theoretical fixpoint theorem and its applications.
\newblock {\em Pacific J. Math.}, 5:285--309, 1955.

\end{thebibliography}

\newpage 
\clearpage
\section{Appendix: complete proofs}

\subsection{On tree with back edges}

\begin{lemma} [i.e. Lemma \ref{root:path}]
  If $\pi$ is a simple path of $b$-length $n$, then $r_{n}$ is the
  vertex closest to the root visited by $\pi$.
\end{lemma}
\begin{proof}
  It is enough to observe that, for each $i$, $r_{i}$ is the highest
  vertex visited by $\pi_{i}$. To this goal, if $\pi_{i} = d_{i} \ast
  b_{i}$, where $d_i$ is a tree path and $b_i$ is a back-edge, then
  either $r_{i}$ belongs to $d_{i}$ or it is an ancestor of the source
  of $d_{i}$. The first case is excluded by $\pi_{i}$ being simple.
  \myqed
\end{proof}

\subsection{A variant of the entanglement game}

\begin{proposition} [i.e. Proposition \ref{modif:entag}]
  Let $\ET{G,k}$ be the game played as the game $\Ent{G,k}$ except
  that Cops is allowed to retire a number of cops placed on the
  graph. That is, Cops moves are of the form
  \begin{itemize}
  \item $(g,C,Cops) \rightarrow (g,C',Thief)$
    % \\ 
    % \mbox{\hspace{20mm}}
    (generalized skip move),
  \item $(g,C,Cops) \rightarrow (g,C'\cup \set{g},Thief)$
    % \\ \mbox{\hspace{20mm}}
    (generalized replace move),
  \end{itemize}
  where in both cases $C' \subseteq C$.
  Then Cops has a winning strategy in $\Ent{G,k}$ if and only of he
  has a winning strategy in $\ET{G,k}$.
\end{proposition}
\begin{proof}
  Since every Cops' move in the game $\Ent{G,k}$ is a Cops' move in
  the game $\ET{G,k}$, and since there is no new kind of moves for
  Thief in the game $\ET{G,k}$, then a Cops' winning strategy in
  $\Ent{G,k}$ can be used to let Cops win in $\ET{G,k}$.

  On the other direction, a winning strategy for Cops in $\ET{G,k}$
  can be mapped to a winning strategy for Cops in $\Ent{G,k}$ as
  follows.

  Each position $(g,C,P)$ of $\Ent{G,k}$ is matched by a position
  $(g,C^{-},P)$ of $\ET{G,k}$ such that $C^{-} \subseteq C$. A Thief's
  move $(g,C,Thief) \rightarrow (g',C,Cops)$ in $\Ent{G,k}$ can
  certainly be simulated by the move $(g,C^{-},Thief) \rightarrow
  (g',C^{-},Cops)$ in $\ET{G,k}$, note that Thief has the ability to
  perform such a move because since if $g' \in C^{-}$ then already $g'
  \in C$.

  Assume that the position $(g,C_0,Cops)$ of $\Ent{G,k}$ is matched by
  the position $(g,C_{0}^{-},Cops)$ of $\ET{G,k}$. From
  $(g,C_{0}^{-},Cops)$, Cops' winning strategy may suggest two kinds
  of moves.
  
  It may suggest a generalized skip $(g,C_{0}^{-},Cops) \rightarrow
  (g,C^{-}_{1},Cops)$ with $C^{-}_{1} \subseteq C^{-}_{0}$.  If this
  is the case, the Cops just skips on from the related position
  $(g,C_0,Cops)$.

  It may suggest a generalized replace move $(g,C^{-}_{0},Cops)
  \rightarrow (g,C^{-}_{1} \cup \set{g},Thief)$.  If $\card{C_0} < k$,
  then the such a move  becomes an add move $(g,C_0,Cops) \rightarrow (g,C_0
  \cup \set{g},Thief)$. Otherwise $\card{C_0} = k$ and $\card{C^{-}_{1}}
  < k$ -- since $g \not \in C^{-}_{1}$ and $\card{C^{-}_{1} \cup
    \set{g}} \leq k$ -- and consequently we can pick $x \in C_0
  \setminus C^{-}_{1}$, this is possible since $C_0 \setminus C^{-}_ 1$ is not empty, because $C^{-}_1 \subseteq C^{-}_{0} \subseteq C_0$ and $\card{C^-_1} < \card{C_0}$.  Observe also that $x \neq g$, since this
  would mean that Thief has been trapped. Therefore the  move  $(g,C^{-}_{0},Cops)
  \rightarrow (g,C^{-}_{1} \cup \set{g},Thief)$ is
  simulated by the replace move $(g,C_0,Cops) \rightarrow
  (g,C_0\setminus \set{x} \cup \set{g},Thief)$. Moreover the invariant $C_1^{-} \cup \set{g} \subseteq C_0 \setminus \set{x} \cup \set{g}$ is maintained.
  \myqed
\end{proof}

\subsection{On the $\star$ property of weak simulations}

\begin{lemma} [i.e. Lemma \ref{LEMMA:uniq:center}]
Let $(R,\varsigma)$ be a
  $\star$-weak simulation of $G$ by $H$.  If $C(h)$ is not empty, then
  there exists an element $c(h) \in V_{G}$ such that for each $(g,g')
  \in C(h)$ either $c(h) = g$ or $c(h) = g'$. If moreover $\card{C(h)}
  \geq 2$, then this element is unique.
\end{lemma}
\begin{proof}
  Clearly the condition holds if $\card{C(h)} \leq 2$, by definition
  \ref{def:starwek}. Let us suppose that $\card{C(h)} \geq 3$.

  Fix two undirected edges $\set{c(h),g_{1}},\set{c(h),g_{2}}$ in the
  undirected version of $C(h)$.  Consider a third undirected edge
  $\set{\tilde{g}_{1},\tilde{g}_{2}} \in C(h)$, so that
  $\card{\set{\tilde{g}_{1},\tilde{g}_{2}} \cup \set{c(h),g_{1}}} =
  3$, and similarly $\card{\set{\tilde{g}_{1},\tilde{g}_{2}} \cup
    \set{c(h),g_{2}}} = 3$.\footnote{%
    Observe that the condition on the cardinality implies that we
    cannot have $(g_{1},g_{2}),(g_{2},g_{1}) \in C(h)$. Thus, the
    requirement that $G$ has no directed cycles of length $2$ is
    somewhat superfluous.  }%
  If $c(h) \notin \set{\tilde{g}_{1},\tilde{g}_{2}}$, then $\set{\tilde{g}_{1},\tilde{g}_{2}} = \set{g_{1},g_{2}}$, thus creating an undirected $3$-cycle and contradicting the condition on the girth of $G$.
  %% 
  % If there is a third undirected edge
  % in the undirected version of $C(h)$ whose both endpoints are
  % different from $c(h)$, then this edge is of the form
  % $\set{g_{1},g_{2}x}$, thus creating a $3$-cycle and contradicting the
  % condition on the girth of $G$.
  \myqed
\end{proof}

\begin{lemma}[i.e. Lemma \ref{LEMMA:SIM:COVER}]
  If $(R,\varsigma)$ is a $\star$-weak simulation of $G$ by $H$ and
  $\rho : K \rTo H$ is a cover, then there exists a $\star$-weak
  simulation $(\tilde{R},\tilde{\varsigma})$ of $G$ by $K$.
\end{lemma}
\begin{proof}
  We construct the $\star$-weak simulation  $(\tilde{R},\tilde{\varsigma})$, where $\tilde{R} \subseteq V_G \times V_K$, as follows 
  $$
  g \tilde{R} k \iff g R \rho (k)
  $$ 
  We consider first $\tilde{R}$ and we prove it to be surjective and
  functional.  Since for each $g \in V_G$ there exists $h\in V_H$ such
  that $g R h$ and since $\rho$ is surjective, then there exists $k
  \in V_K$ such that $h=\rho(k)$, and hence $g R \rho(k)$, thus $g
  \tilde{R} k$. Therefore $\tilde{R}$ is surjective. \newline If $g_i
  \tilde{R} k$, $i=1,2$, then $g_i R \rho(k)$. Since $R$ is
  functional, then $g_1=g_2$. Therefore $\tilde{R}$ is
  functional.\newline We exhibit $\tilde{\varsigma}$ as follows. If $g
  \tilde{R} k_0$ and $g\to g'$, then, we take
  $\tilde{\varsigma}(g,g',k_0)=k_0,\dots,k_n$, such that
  $\varsigma(g,g',\rho(k_0))=\rho(k_0),\dots, \rho(k_n)$. Note that
  the path $k_0,\dots,k_n$ is unique. Therefore,
  $(\tilde{R},\tilde{\varsigma})$ is a weak simulation.

  Finally, whenever $(g,g'),(\tilde{g},\tilde{g}')$ are distinct edges
  of $G$ and %\\
  $k_i \in \tilde{\varsigma} (g,g',k_0)\cap \tilde{\varsigma}
  (\tilde{g},\tilde{g}',k_0)$, then $\rho(k_i) \in \varsigma
  (g,g',\rho(k_0)) \cap \varsigma (\tilde{g},\tilde{g}',\rho(k_0))$.
  Since $(R,\varsigma)$ has the $\star$-property, we get
  $|\set{g,g',\tilde{g},\tilde{g}'}|=3$.  It follows that
  $(\tilde{R},\tilde{\varsigma})$ has the $\star$-property.  \myqed
\end{proof}

\subsection{Properties of strongly synchronizing games}

\begin{lemma}
  If $G$ is strongly synchronizing, then the unique winning strategy
  in the game $\homl G,G \homr$ is the copycat strategy.
\end{lemma}
\begin{proof}
  Let us consider a position $g \in Pos^{G}_{E}$, and let us analyze
  the position $(g,g)$ of $\homl G,G \homr$. Let us suppose that $(g,g') \in M^G$ and consider the possible Mediator's answers to the
  Opponents' move $(g,g) \rightarrow (g',g)$.

  Mediator cannot answer $(g',g) \rightarrow (g'',g)$, since then the
  relation $(G,g'') \leq (G,g)$ implies that either $g'' = g$ (hence
  having a cycle of length $2$ in $G$), or that there is an undirected
  edge between $g''$ and $g$, thus creating a length $3$ cycle.

  Similarly Mediator cannot answer $(g',g) \rightarrow (g',\tilde{g})$
  with $g' \neq \tilde{g}$. Again, this would create a length $3$
  cycle in the undirected version of $G$.
  \myqed
\end{proof}

\begin{lemma} [i.e. lemma \ref{Lemma:Intersec}] %
 % \label{intersection1:lemma}% 
 % \label{intersection2:lemma}%
  Let $G$ be a strongly synchronizing  and
  $(g,g'),(\tilde{g}, \tilde{g}') \in M^G$.
  \begin{enumerate}
  \item   If $(G,g) \sim \hat{x}$ then $g \in
    Pos^{G}_{D}$ and $\lambda(g) = x$.
  \item If $g,\tilde{g} \in Pos_{E}^G$ and,
    for some game $H$ and $h \in Pos^{H}$, we have
    \begin{align*}
      (G,g') & \leq (H,h) \leq (G,g) 
      %&%
      \tand %&
      %&%
      (G,\tilde{g}') %& 
      \leq (H,h) \leq (G,\tilde{g})\,,
    \end{align*}   
    then $g=\tilde{g}$ or $g'=\tilde{g}'$, and
    $|\set{g,g',\tilde{g},\tilde{g}'}|=3$.
  \item If $g \in Pos_{E}^G$ and $\tilde{g} \in Pos_{A}^G$ and, for some
    $H$ and $h \in Pos^H$, we have
    \begin{align*}
      (G,g') & \leq (H,h) \leq (G,g) 
      %&
      %% 
      \tand 
      %&
      %%
      %&
      (G,\tilde{g}) %& 
      \leq (H,h) \leq (G,\tilde{g}')\,,
    \end{align*}
    then $g=\tilde{g}'$ or $g'=\tilde{g}$, and
    $|\set{g,g',\tilde{g},\tilde{g}'}|=3$.
  \end{enumerate}
\end{lemma}

\begin{proof}
\begin{enumerate} 
\item 
%\spnewtheorem*{fact}{Fact}{\bfseries}{\itshape}
Let $\chi_{G}$ be the set of free variables of $G$. First, we have the following claim.
\begin{claim}
If $(G,g) \sim \hat{x}$, then $x\in \chi_{G}$.
\end{claim}
\begin{proof}
On the one hand, if $x \notin \chi_G$ then $G[x/ \top]\sim G[x/ \bot]$. One the other hand, $G[x/ \top] \sim \hat{x} [x /\top] \sim \top$ and $G[x/ \bot] \sim \hat{x} [x /\bot] \sim \bot$, thus $\bot =\top$. This ends the proof of the claim.\myqed
\end{proof}
If $g$ has a successor, then the winning strategy in $\homl G,\hat{x},G \homr$  will suggest for example to play $(g,p_{\star}^{\hat x},g) \to (g',p_{\star}^{\hat x},g) \to (g',p_{\star}^{\hat x},g')$, for some $(g,g') \in M^G$. Therefore $(G,g) \sim\hat{x} \sim (G,g')$, contradicting the fact that $G$ is strongly synchronizing. Thus $g$ has no successor, and clearly $g \in Pos_{D}^G$ and  $\lambda^{G}(g)=x$, according to the claim.

\item  We derive first $(G,g') \leq (G,\tilde{g})$ and $(G,\tilde{g}') \leq (G,g)$ and
  observe that each inequality is strict, because the game is bipartite.
% i.e. for example, if $(G,g')=(G,\tilde{g})$ then $g' = \tilde{g}$, hence $g \rightarrow \tilde{g}$ with $\lambda(g) = \lambda(\tilde{g})$.
  Therefore from item 2 of Definition \ref{StrongSynch} we have a
  diagram of the form
  $$
  \xygraph{
    []*+{g}="g"
    :[d]*+{g'}
    -[r]*+{\tilde{g}}="tg"^{<}
    :[u]*+{\tilde{g}'}="tgp"
    -"g"_{>}
  }\,
  $$ 
  that is we have an undirected edge bewteen $g$ and $\tilde{g}'$, and
  an undirected edge between $g'$ and $\tilde{g}$.
  
  If $g\neq\tilde{g}$ and $g'\neq\tilde{g}'$, then the above diagram
  gives rise to an undirected cycle of length $4$, which cannot
  happen.
\item As before, we derive $(G,\tilde{g}) \leq (G,g)$ and $(G,g') \leq
  (G,\tilde{g}')$ and moreover $(G,\tilde{g}) < (G,g)$ and $(G,g') <
  (G,\tilde{g}')$, since $g$ and $\tilde{g}$ belong to opposite
  players. Therefore from item 2 of definition \ref{StrongSynch} we
  obtain a diagram of the form
  $$
  \xygraph{
    []*+{g}="g"
    :[d]*+{g'}
    -[r]*+{\tilde{g}'}="tgp"^{<}
    [u]*+{\tilde{g}}
    (
    :"tgp",-"g"_{>}
    )
  }
  $$
  If $g\neq\tilde{g}'$ and $g'\neq\tilde{g}$, then the above diagram
  gives rise to an undirected cycle of length $4$, which cannot
  happen. 
\end{enumerate}
\myqed
\end{proof}
%--------------------------------------------------------

\subsection{The games $G_{n}$ are strongly synchronizing}
It is clear that the game $G_n$ is bipartite and $\Ent{G_n}=n$,
moreover the girth of $G_n$ is $6$. To accomplish the proof that $G_n$
is stronlgy synchronizing, we need some intermediary lemmas.
%\spnewtheorem*{claim}{claim}{\bfseries}{\itshape}

% To make sense of the the next Lemmas and Propositions, let
% \begin{align*}
%   In_{n} & = \set{ (i,j,k) \in [n] \times [n] \times [k]\mid j = 0
%     \tiff k = 0}\,.
% \end{align*}
% Let us identify $v_{i}$ with $v_{i,0,0}$ and 
% $w_{i}$ with $w_{i,0,0}$.

\begin{lemma}   \label{lemma:zerosigma}
  If $(G_n,w_{i,j,k}) \leq (G_n,g)$ then either $g = w_{i,j,k}$ or
  $g \in Pos_{E}^{G_n}$ and $g = v_{i,j,k}$.
\end{lemma}
\begin{proof}
  \begin{proofbycases}
    \begin{caseinproof}
      If $g = w_{i',j',k'}$, then surely we need to have $(i,j,k) =
      (i',j',k')$.
    \end{caseinproof}
  Let therefore $g = v_{i',j',k'}$.
  \begin{caseinproof}
    \label{case:secondcase}
    If $g\in Pos_{A}^{G_n}$ and $(i,j,k) \neq (i',j',k')$, Opponents can
    choose to move $(w_{i,j,k},v_{i',j',k'}) \rightarrow
    (w_{i,j,k},w_{i',j',k'})$, the latter being a lost position for
    Mediator.
  \end{caseinproof}  %%
  \begin{caseinproof}
    \label{case:thirdcase}
    If $g \in Pos_{A}^{G_n}$ and $(i,j,k) = (i',j',k')$, Opponents can
    choose to move $(w_{i,j,k},v_{i,j,k}) \rightarrow
    (w_{i,j,k},v_{i',j',k'})$ with $(i,j,k) \neq (i',j',k')$. From
    this position Mediator cannot move $(w_{i,j,k},v_{i',j',k'})
    \rightarrow (w_{i,j,k},w_{i',j',k'})$, nor
    $(w_{i,j,k},v_{i',j',k'}) \rightarrow
    (w_{i,j,k},v_{i'',j''',k''})$, since the girth of $G_{n}$ being
    equal to $6$ implies that $(i,j,k) \neq (i'',j'',k'')$ and
    $v_{i'',j'',k''} \in Pos_{A}^{G_n}$, falling back into case
    \ref{case:secondcase}.
  \end{caseinproof}  %%
  \begin{caseinproof}
    If $g \in Pos_{E}^{G_n} $ and $(i,j,k) \neq (i',j',k')$, then
    Mediator cannot move $(w_{i,j,k},v_{i',j',k'}) \rightarrow
    (w_{i,j,k},w_{i',j',k'})$.  He cannot either move
    $(w_{i,j,k},v_{i',j',k'}) \rightarrow (w_{i,j,k},v_{i'',j'',k''})$
    since $v_{i'',j'',k''} \in Pos_{A}^{G_n}$, thus falling back 
    either into case \ref{case:secondcase}, or into case
    \ref{case:thirdcase}.
  \end{caseinproof}
  Therefore, the only possibility is that 
  $g \in Pos_{E}^{G_n}$ and $(i,j,k) = (i',j',k')$.
  \end{proofbycases}
  \myqed
\end{proof}
Dualizing the previous proof we obtain:
\begin{lemma}  \label{lemma:pizero}
  If $(G_n,g) \leq (G_n,w_{i,j,k})$ then either $g = w_{i,j,k}$ or
  $g \in Pos_{A}^{G_n}$ and $g = v_{i,j,k}$.
\end{lemma}

%% What is the need of that ???
% And moreover:
% \begin{lemma} \label{lemma:pizero2}
%  If $(G_n,g) \leq (G_n,w_i)$ then either $g = w_{i}$ or $g=v_i$.
% \end{lemma} 

\begin{lemma}  \label{lemma:pisigma}
  If $(G,v_{i,j,k}) \leq (G,v_{i',j',k'})$ and $v_{i,j,k} \neq
  v_{i',j',k'}$, then either $v_{i,j,k} \in Pos_{A}^{G_n}$ and $(v_{i,j,k}
  , v_{i',j',k'}) \in M^{G_n}$, or $v_{i',j',k'} \in Pos_{E}^{G_n}$ and
  $(v_{i',j',k'}, v_{i,j,k}) \in M^{G_n}$.
\end{lemma}
\begin{proof}
  \begin{proofbycases}
    Let us suppose that $v_{i,j,k} \in Pos_{A}^{G_n}$. We remark that
    $v_{i',j',k'} \not\in Pos^{G_{n}}_{D}$, and thus we split the
    proof into two cases.
    %\begin{caseinproof}
    %  We observe first that by Lemma \ref{lemma:pizero} we cannot have
    %  $\lambda(v_{i',j',k'}) = 0$.
    %\end{caseinproof}  %%
    %% 
    \begin{caseinproof}
      \label{case:previous}
      If $v_{i',j',k'} \in Pos_{A}^{G_n}$, then Opponents can move
      $(v_{i,j,k},v_{i',j',k'}) \rightarrow (v_{i,j,k},w_{i',j',k'})$.
      This is a lost position  by Lemma \ref{lemma:pizero}.
    \end{caseinproof}  %%
    \begin{caseinproof}
      Therefore we have $v_{i',j',k'} \in Pos_{E}^{G_n}$.  Mediator
      has two kinds of moves. He can choose to move to a ``variable'',
      that is, to move $(v_{i,j,k},v_{i',j',k'}) \rightarrow
      (v_{i,j,k},w_{i',j',k'})$ or $(v_{i,j,k},v_{i',j',k'})
      \rightarrow (w_{i,j,k},v_{i',j',k'})$. These moves, however,
      lead to lost positions, by Lemmas \ref{lemma:zerosigma} and 
      \ref{lemma:pizero}.  
      %% Similarly, he cannot choose a variable on
      %% the left.
      %% 
      Therefore, if the position $(v_{i,j,k},v_{i',j',k'})$ is
      winning, then he can only move $(v_{i,j,k},v_{i',j',k'})
      \rightarrow (v_{i,j,k},v_{i'',j'',k''})$ or
      $(v_{i,j,k},v_{i',j',k'}) \rightarrow
      (v_{i'',j'',k''},v_{i',j',k'})$.  In the first case, if the
      position $(v_{i,j,k},v_{i'',j'',k''})$ is winning, then $(i,j,k)
      = (i'',j'',k'')$  by case \ref{case:previous}; hence
      $(v_{i',j',k'} , v_{i,j,k}) \in M^{G_n}$.
      In the second case, if Mediator moves to a winning position
      $(v_{i,j,k},v_{i',j',k'}) \rightarrow
      (v_{i'',j'',k''},v_{i',j',k'})$, then $(i',j',k') =
      (i'',j'',k'')$ by the dual of case \ref{case:previous} and hence
      $(v_{i,j,k}, v_{i',j',k'}) \in M^{G_n}$.
    \end{caseinproof}
  \end{proofbycases}
\end{proof}
%-----------------------
Thus we are ready to prove:
\begin{proposition}
  The games $G_n$ are strongly synchronizing.
\end{proposition}
\begin{proof}
  Let us prove first that $(G,g) \sim (G,\tilde{g})$ implies $g =
  \tilde{g}$.
  \begin{proofbycases}
    Let us assume that $(G,g) \sim (G,\tilde{g})$, we split the proof
    that $g = \tilde{g}$ into three cases, according to the color of
    $g$.
    \begin{caseinproof}
      Assume $g \in Pos_{D}^{G_n}$ and thus let $g = w_{i,j,k}$.  If
      $g \neq \tilde{g}$, then Lemma \ref{lemma:zerosigma} implies
      that $\tilde{g} = v_{i,j,k}$ with $\tilde{g} \in Pos_{E}^{G_n}$.
      Similarly Lemma \ref{lemma:pizero} implies that $\tilde{g} =
      v_{i,j,k}$ with $\tilde{g} \in Pos_{A}^{G_n}$.  Thus we reach a
      contradiction, and therefore $g = \tilde{g}$.
    \end{caseinproof}
    \begin{caseinproof}
      Let us assume that $g = v_{i,j,k} \in Pos_{E}^{G_n}$. Then
      $(G,w_{i,j,k}) < (G,g) \sim (G,\tilde{g})$ and therefore
      $\tilde{g} = v_{i,j,k}$ by Lemma \ref{lemma:zerosigma}.
    \end{caseinproof}
    \begin{caseinproof}
      If $g = v_{i,j,k}\in Pos_{A}^{G_n}$ then $(G,\tilde{g}) \sim
      (G,g) < (G,w_{i,j,k})$ and therefore $\tilde{g} = v_{i,j,k}$ by
      Lemma \ref{lemma:pizero}.
    \end{caseinproof}
  \end{proofbycases}

  \breath

  Let us now prove that $(G,g) \leq (G,\tilde{g})$ and $g \neq
  \tilde{g}$ implies $\tilde{g}\in Pos_{E}^{G_n} \tand (\tilde{g},g)
  \in M^{G_n}$ or $g\in Pos_{E}^{G_n} \tand (g,\tilde{g}) \in
  M^{G_n}$.
  
  This is the case if $g \in Pos_{D}^{G_n}$ or $\tilde{g} \in
  Pos_{D}^{G_n}$, by Lemmas \ref{lemma:zerosigma} and
  \ref{lemma:pizero}. If both $g,\tilde{g} \in Pos_{E,A}^{G_n}$, then
  the statement follows from Lemma \ref{lemma:pisigma}.  \myqed
\end{proof}

%%% Local Variables: 
%%% mode: latex
%%% TeX-master: "0"
%%% End: 

\end{document}